\pdfoutput=1
\documentclass[a4paper,fleqn, usenatbib]{mnras}

\usepackage{newtxtext,newtxmath}
\usepackage[T1]{fontenc}
\usepackage{ae,aecompl}

\usepackage{graphicx}	
\usepackage{amsmath}	
\usepackage{amssymb}	

\title[Dust and star formation profiles]{\emph{UVI} colour gradients of $0.4<\lowercase{z}<1.4$ star-forming main sequence galaxies in CANDELS: dust extinction and star formation profiles}

\author[W. Wang et al.]{Weichen Wang$^{1,2}$\thanks{E-mail: wcwang@jhu.edu},
	S. M. Faber$^{3}$, 
	F. S. Liu$^{4}$,
	Yicheng Guo$^{3}$,
	Camilla Pacifici$^{5}$\thanks{NASA Postdoctoral Program Fellow}, \newauthor 
	David C. Koo$^{3}$,
	Susan A. Kassin$^{6}$, 
	Shude Mao$^{1,7,8}$,
	Jerome J. Fang$^{9}$,
	Zhu Chen$^{10}$,  \newauthor
	Anton M. Koekemoer$^{6}$,
	Dale D. Kocevski$^{11}$, 
	M. L. N. Ashby$^{12}$	
\\
$^{1}$Department of Physics and Tsinghua Center for Astrophysics, Tsinghua University, Beijing 100084, China \\
$^{2}$Department of Physics \& Astronomy, Johns Hopkins University, 3400 N. Charles Street, Baltimore, MD 21218, USA\\
$^{3}$Department of Astronomy and Astrophysics, University of California Observatories, University of California, Santa Cruz, CA 95064, USA\\
$^{4}$College of Physical Science and Technology, Shenyang Normal University, Shenyang 110034, China\\
$^{5}$NASA Goddard Space Flight Center, Code 665, Greenbelt, MD, USA\\
$^{6}$Space Telescope Science Institute, 3700 San Martin Drive, Baltimore, MD 21218, USA\\
$^{7}$National Astronomical Observatories, Chinese Academy of Sciences, 20A Datun Road, Beijing, 100012, China\\
$^{8}$Jodrell Bank Centre for Astrophysics, University of Manchester, Alan Turing Building, Manchester M13 9PL, UK\\
$^{9}$Orange Coast College, Costa Mesa, CA 92626, USA\\
$^{10}$Shanghai Normal University, 100 Guilin Road, Shanghai, 200234, China\\
$^{11}$Department of Physics and Astronomy, Colby College, Waterville, ME 04961, USA\\
$^{12}$Harvard-Smithsonian Center for Astrophysics, 60 Garden St., Cambridge, MA 02138, USA
}
\date{Accepted XXX. Received YYY; in original form ZZZ}

\pubyear{2017}
\begin{document}
\label{firstpage}
\pagerange{\pageref{firstpage}--\pageref{lastpage}}
\maketitle
	\begin{abstract}
		This paper uses radial colour profiles to infer the distributions of dust, gas and star formation in $z=0.4$--1.4 star-forming main sequence galaxies.  We start with the standard \textit{UVJ}-based method to estimate dust extinction and specific star formation rate (sSFR).  By replacing $J$ with $I$ band, a new calibration method suitable for use with ACS+WFC3 data is created (i.e. \textit{UVI} diagram).  Using a multi-wavelength multi-aperture photometry catalogue based on CANDELS, UVI colour profiles of 1328 galaxies are stacked in stellar mass and redshift bins. The resulting colour gradients, covering a radial range of 0.2--2.0 effective radii, increase strongly with galaxy mass and with global $A_V$.  Colour gradient directions are nearly parallel to the Calzetti extinction vector, indicating that dust plays a more important role than stellar population variations.  With our calibration, the resulting $A_V$  profiles fall much more slowly than stellar mass profiles over the measured radial range. sSFR gradients are nearly flat without central quenching signatures, except for $M_\star>10^{10.5} \mathrm{M}_{\sun}$, where central declines of 20--25 per cent are observed.  Both sets of profiles agree well with previous radial sSFR and (continuum) $A_V$ measurements. They are also consistent with the sSFR profiles and, if assuming a radially constant gas-to-dust ratio, gas profiles in recent hydrodynamic models.  We finally discuss the striking findings that SFR scales with stellar mass density in the inner parts of galaxies, and that dust content is high in the outer parts despite low stellar-mass surface densities there.  
	\end{abstract}
	
	
	\section{Introduction} 
	\label{sec:intro}
	Radial star-formation gradients are crucial for determining how stellar
	mass is built up in galaxies as they evolve along the star-forming main 
	sequence (e.g.,  \citealt{Brinchmann2004, Noeske2007, Elbaz2007}), and how and where the star-forming process shuts 
	down. UV-optical star formation
	measurements are typically ambiguous without proper
	dust corrections (e.g., \citealt{Meurer1999, Kriek2009, Brammer2009, Patel2011b, Murphy2011, Reddy2012}), so 
	spatially resolved star-formation and dust need to be measured 
	together.   Moreover, a significant fraction of high-redshift star-forming 
	galaxies are very
	dusty, with integrated $A_V$ values up to 2.5 mag
	(\citealt{Williams2009, Brammer2009, Patel2011b}). This
	means that dust could be seriously distorting our view of the 
	structure of these 
	galaxies as observed  in
	optical light.  
	Dust gradients have often been overlooked or treated to be  subordinate when
	studying distant galaxies (e.g., \citealt{Tacchella2015, Nelson2016a}), 
	which may cause non-negligible biases in conclusions about the SFR and stellar age distributions. 
	
	Radial $A_V$ profiles may also give insights into the
	gas surface density distribution, if provided with the inference of gas-to-dust ratio (e.g.\ \citealt{Rachford2009, Bolatto2013, Sandstrom2013, Genzel2013} ).
	Measuring gas distributions in distant galaxies requires high angular
	resolution, which is not usually available at radio wavelengths.  
	$A_V$ maps can substitute for gas maps until better data are available.
	
	Only a few methods have been used so far to measure radial dust
	and star-formation rate (SFR) gradients
	in distant galaxies.  The main challenge is obtaining high-resolution images
	at all wavelengths that each method requires.
	H\,$\alpha$ and H\,$\beta$ images from 3D-HST  grism data
	\citep{Brammer2012, Momcheva2016} yield maps of H\,$\alpha$ 
	and the Balmer decrement
	as a function of position \citep{Nelson2016}. However, the S/N is 
	low, and samples have to be extensively stacked in order to obtain results.
	Moreover, using  the
	popular WFC3/G141 grism, both emission lines are only available in the 
	limited redshift range
	$z=1.35$--1.5. The UV $\beta$-slope method \citep{Meurer1995, Meurer1997,Meurer1999} measures
	attenuation from the excess
	slope of the far-UV stellar continuum but is reliable only for objects having long UV exposures, because ultraviolet photons are heavily absorbed.
	\textit{ALMA} (the Atacama Large Millimeter/submillimeter Array) is providing new 
	high-resolution information in the form of images of cool dust shrouding 
	star-forming regions (e.g., \citealt{Barro2016a, Dunlop2016}).
	However, the current \textit{ALMA} sample pool is not large enough for statistical 
	studies.
	
	The most common method for dust determination at high $z$
	involves fitting attenuated stellar population models to the \emph{integrated}
	broad-band spectral energy distributions of galaxies (e.g., 
	\citealt{Kriek2009}).
	\citet{Patel2011b} show that this method is closely
	related to the classic $U-V$ versus $V-J$ two-colour method (\textit{UVJ}) that
	distinguishes dust reddening from old stars.  The so-called \textit{UVJ} diagram 
	has been widely utilized to identify quenched galaxies \citep{Labbe2005, Wuyts2007, Williams2009} and also to give $A_V$ and 
	the dust-corrected
	specific star formation rate (sSFR) values consistent with the
	conventional SED fitting results Fang et al.~(2017, submitted).
	Broad-band \emph{HST} images also have the advantages of high S/N,
	wide accessible redshift range,
	and large samples to permit investigating second-parameter effects.

	As of this writing, only one paper, \citet{Liu2016}, has
	used high-resolution \emph{HST} optical-IR images (in $B$ through $H$ band) 
	to measure dust and
	sSFR gradients in star-forming main sequence galaxies
	(hereafter SFMS galaxies).   Their analysis uses an annular aperture 
	photometric catalogue
	based on images taken as part of the CANDELS program (Cosmic Assembly Near-infrared Deep 
	Extragalactic
	Legacy Survey, \citealt{Grogin2011, Koekemoer2011}) on 
	\emph{HST}.
	The study concluded that at $z=0.4$--1.4, dust extinction is the principal 
	cause of rest-frame $NUV-B$ colour gradients at low stellar mass ($M_\star 
	< 10^{10.5} \mathrm{M}_{\sun}$),
	while for high-mass galaxies ($M_\star>10^{10.5} \mathrm{M}_{\sun}$) age gradients are also 
	a factor.
	
	The present paper is built upon the same annular photometric catalogue used by
	\citet{Liu2016} but with two differences.
	First, \citet{Liu2016} used the \textsc{fast} spectral fitting program \citep{Kriek2009}  
	to obtain
	$A_V$ and sSFR in each galaxy annulus.  We instead base our results
	on the same \textit{UVJ} method as used by Fang et al.~(2017, submitted), which has the advantage of showing the raw
	data directly on the \textit{UVJ} plane and providing a simple intuitive geometric
	interpretation of them.
	Second, the SED fits of \citet{Liu2016} lacked long-wavelength (IRAC)
	data points because high-resolution \emph{HST} imaging 
	ends at $H$ band.  Because the effects of omitting IRAC from SED fitting
	are not well known, we prefer to use a method that is
	more securely based on IRAC data.  The \textit{UVJ} calibration 
	based on integrated galaxy photometry including IRAC, meets this need as well.
	
	However, reproducing \textit{UVJ} requires an indirect approach for $z\sim 1$
	because rest-frame $J$ band has shifted to the observed $H$ band even at $z = 0.4$.
	We therefore first verify that $U-V$ versus $V-I$ diagram (hereafter $UVI$) works nearly as well as \textit{UVJ} for deriving
	$A_V$ and sSFR and then
	synthesize usable $V-I$ values from available \emph{HST} $B$ through $H$
	imaging. This  proves to be
	possible all the way out to $z=1.4$ because of the simple, log-linear 
	slope of galaxy
	continua near 10,000 \AA, which generally lack strong spectral 
	features (except for emission in strongly star-forming galaxies; see Appendix \ref{append:A}).  The \textit{UVJ} 
	calibrations from Fang et al.~(2017, submitted) for $A_V$ and sSFR are then 
	transferred to the \textit{UVI} plane
	in a manner that ensures no bias.
	
	The organization of this paper is as follows.
	First we stack raw galaxy colour gradients in stellar mass 
	and redshift bins and study their behavior on the \textit{UVI} 
	diagram. Basic features such as the raw colour-gradient directions and 
	correlations with mass and redshift are revealed.
	The \textit{UVJ}-based calibrations are then utilized to convert raw \textit{UVI} colours 
	into values of $A_V$ and sSFR, which are used to address the following 
	questions:
	\begin{enumerate}
	\item  What is the average $A_V$ radial profile and how does 
	it vary with
	stellar mass, redshift, and inclination?
	How does dust surface density vary radially in comparison to stellar mass?
	
	\item  What do dust-corrected sSFR profiles look like? Are there signs
	of bulge quenching at small radii?  How are these results affected by
	uncertainties in the stellar population
	models and the attenuation law?
	
	\item  How do our results compare to previous measurements
	and to recent hydrodynamic models of galaxy formation?
	\end{enumerate}
	The basic calibrations for both $A_V$ and sSFR in this work
	were originally developed by Fang et al.~(2017, submitted).  The $A_V$ values used are based
	on average fits by different authors to CANDELS photometry, which have been tabulated 
	in \citet{Santini2015}.
	They amount to running SED fitting codes (like \textsc{fast}, \citealt{Kriek2009}) 
	on integrated galaxy
	SEDs from $B$ through IRAC passbands. The sSFR values used Fang et al.~(2017, submitted) technically 
	differ from the
	SED fitting values, but the differences are small and can be ignored.
	The SED fits assume declining
	$\tau$-models and a foreground-screen Calzetti attenuation law 
	\citep{Calzetti1994,Calzetti2000}, 
	and most assume solar-metallicity. We collectively refer to this 
	SED-fitted body of data
	as the `conventional method' for finding $A_V$ and sSFR.  It is possible 
	that some of our
	conclusions depend on this particular recipe, and a critical discussion 
	is given in Section \ref{sec:discussion}.

	A flat $\Lambda$CDM cosmology model with $\Omega_{M}=0.3$ and 
	$\Omega_{\Lambda}=0.7$ is adopted throughout this paper, and the Hubble 
	constant is set to be 70 km s$^{-1}$ Mpc$^{-1}$.  The letter $r$ 
	represents the radial distance to galaxy centres along the semi-major axis unless otherwise 
	stated. \textit{SMA}  represents the size of the effective radius along the 
	galaxy semi-major axis.
	
	\begin{table*}
		\caption{Sample Selection Cuts in the GOODS-South and UDS Fields}
		\label{tab:sample}
		\centering 
		\begin{tabular}{cccc}
		\hline 
		Cut & GOODS-South & UDS & Total\\	
		\hline
		Full catalogue& 34930 (100\%)&	35922 (100\%)	& 70852 (100\%) \\	
		$m_{\mathrm{WFC3/F160W}}\leqslant 24.5$&8565 (24.52\%)&	10077 (28.05\%)&	18642 (26.31\%) \\
		$\mathrm{CLASS\_STAR}<0.9$& 8307 (23.78\%)&	9686 (26.96\%)&	17993 (25.40\%) \\
		Photometry $\mathrm{FLAGS}=0$& 8091 (23.16\%)&	9012 (25.09\%)&	17103 (24.14\%) \\
		$0.4<z<1.4$& 4188 (11.99\%)&	4671 (13.00\%)&	8859 (12.50\%) \\
		$9.0<\log M_{\star}/\mathrm{M}_{\sun}<11.0$& 2477 (7.09\%)&	2725 (7.59\%)&	5202 (7.34\%) \\
		\textsc{galfit} $\mathrm{FLAG}=0$& 2204 (6.31\%)&	2505 (6.97\%)&	4709 (6.65\%) \\
		Aperture photometry available& 1984 (5.68\%)&	1895 (5.28\%)&	3879 (5.47\%) \\
		$\Delta \mathrm{sSFR}>-0.45$ dex & 1513 (4.33\%)&	1428 (3.98\%)&	2941 (4.15\%) \\
		$b/a>0.5$& 797 (2.28\%)&	773 (2.15\%)&	1570 (2.22\%) \\
		$r_{e, \mathrm{F160W}}>0.18\arcsec$& 670 (1.92\%) & 658 (1.83\%)&	1328 (1.87\%) \\
		\hline 
		
		\end{tabular}
	\end{table*}
	
	\begin{figure}
		\centering
		\includegraphics[width=\columnwidth]{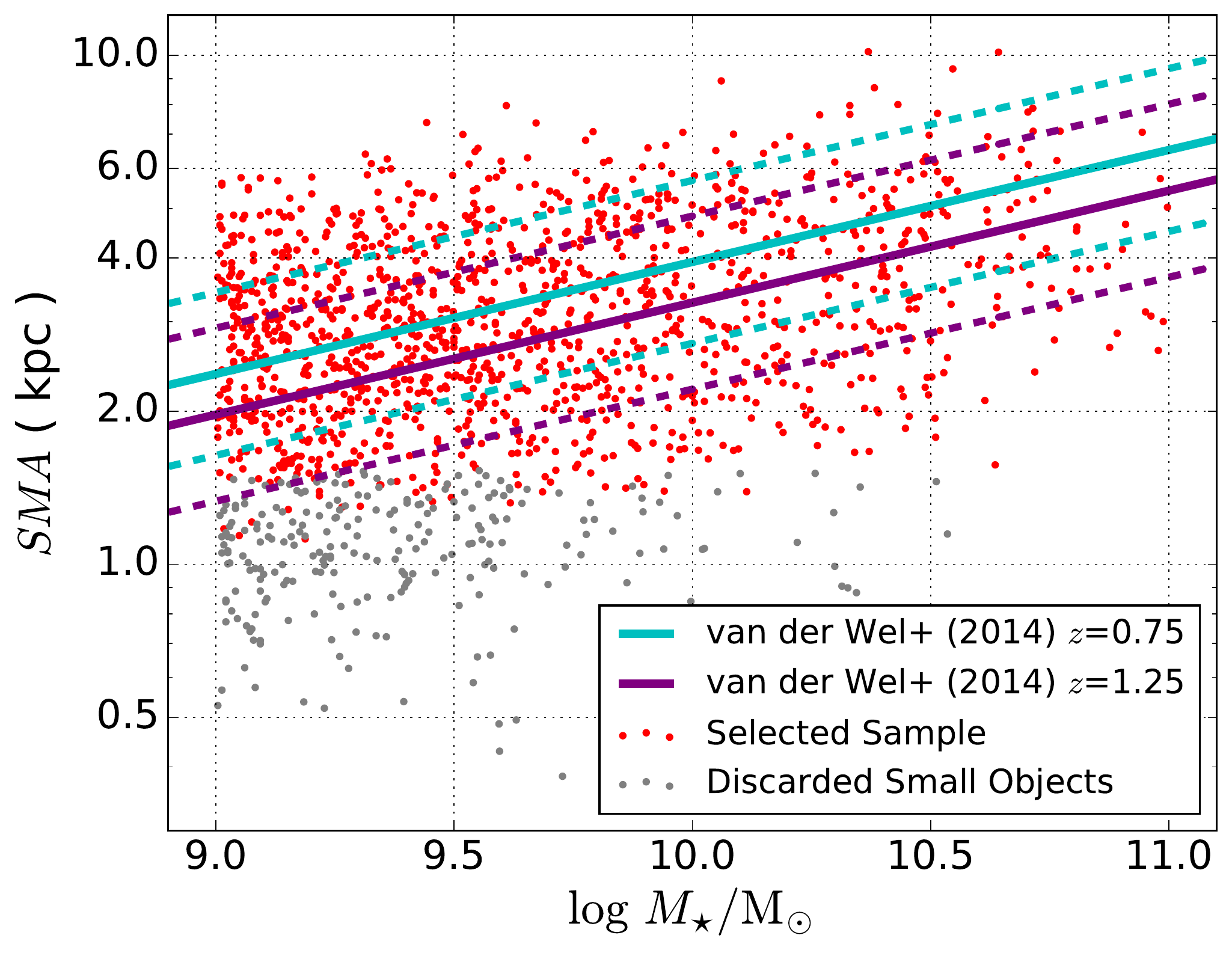}
		\caption{Distribution of \textit{SMA} (effective radius along the galaxy semi-major axis), measured in the WFC3/F160W band, versus stellar mass of the sample galaxies. Red points represent the selected objects; gray points are those discarded after the angular size cut (\textit{SMA}$>0.18$\arcsec). Loss due to small angular size is larger at small masses. Solid colour lines stand for the galaxy size-mass
		relation along the SFMS at around $z=0.75$ and $z=1.25$ measured by \citet{VanderWel2014a} at rest-frame 5000 \AA, with $1\sigma$ scatter indicated by the dashed lines. Objects discarded due to small angular size lie among in the lower outliers of the size-mass relation and mostly beyond the $1\sigma$ scatter range. The fraction of gray points becomes significant only for $\log M_\star/\mathrm{M}_{\sun}<9.7$.}
		\label{fig:sample}
	\end{figure}
	
	\section{Data and Sample Selection} 
	\label{sec:data}
	
	Our work is based on two of the five CANDELS fields, GOODS-South and UDS, for which photometry and galaxy physical parameters have been well studied and compiled by the CANDELS team. We make use of the multi-wavelength photometry catalogues in GOODS-S and UDS \citep{Guo2013, Galametz2013}, the corresponding physical parameter catalogue  \citep{Santini2015}, the morphology structure catalogue \citep{VanderWel2012}, the SFR and $A_V$ catalogue constructed by Fang et al.~(2017, submitted), and the multi-aperture photometry catalogue currently under construction by Liu et al. (in prep.; see \citealt{Liu2016} for more details). 
	
	Photometry measurements of the two fields are provided by \citet{Guo2013} and \citet{Galametz2013}. PSF-matching has been applied to these multi-wavelength photometry  data using the \textsc{tfit} routine \citep{Laidler2007}.  Redshifts are obtained from a joint database of photometric redshifts and spectroscopy redshifts. Spectroscopy redshifts \citep{Morris2015} are adopted in preference if available, and the median photometric redshifts deduced from 11 different photo-$z$ fitting methods are used otherwise \citep{Dahlen2013}.  The redshift values are used as input to obtain integrated rest-frame magnitudes from $FUV$ to $K$ band using the \textsc{eazy} code \citep{Brammer2008} and the templates from \citet{Muzzin2013}, as implemented by the CANDELS team collaboration  (D. Kocevski, S. Wuyts and G. Barro).  \textsc{galfit} fitting results \citep{Peng2002a} are available in the WFC3/F160W band, including the axis ratio $b/a$, S\'ersic index from the single-component fitting, and the half-light radius along the semi-major axis (\textit{SMA}) \citep{VanderWel2012}. 
	
	For stellar masses, we adopt the CANDELS `reference' mass, which is the median of all SED fitting results after being scaled to the Chabrier IMF \citep{Chabrier2003}, to lessen potential biases caused by any specific choice of SED fitting assumptions \citep{Santini2015}. The stellar mass obtained has a nominal scatter of around 0.1 dex.  See \citet{Santini2015} for details. 
	
	The galaxy-integrated sSFR and $A_V$ values are adopted from Fang et al.~(2017, submitted). Global visual dust attenuation $A_{V}$  is the median value of five different modeling results from the physical parameter catalogue \citep{Santini2015}, 2a$_\tau$, 2d$_\tau$, 12a$_\tau$, 13a$_\tau$, and 14a$_\tau$.  These all follow the conventional model, i.e., an exponentially declining star formation history (SFH) and the Calzetti dust extinction law applied as foreground screen. Three of the five methods have fixed stellar metallicity to solar abundance (2a$_\tau$, 13a$_\tau$, 14a$_\tau$). Among the five methods, the 2a$_\tau$ value is seen to resemble most closely to the median $A_V$ value, with the smallest scatter of $0.25$ mag and no systematic deviation detected. Therefore we equate the median $A_V$ values used in this work as an equivalent outcome of solar metallicity SED models. 
	
	Star formation rates (SFR) are the same as derived by Fang et al.~(2017, submitted) 
	based on the luminosity at 2800 \AA, after the correction of dust extinction at 2800 \AA \ \citep{Calzetti2000, KennicuttJr2012}:
	\begin{equation}
		\mathrm{SFR}[\mathrm{M}_{\sun} yr^{-1}]= 2.59\times 10^{-10} L_{NUV, \mathrm{ corrected} }[\mathrm{L}_{\sun}]
	\end{equation}
	
	Fang et al.~(2017, submitted)
	tested the robustness of this method by comparing the sSFR values with those derived from UV and far-IR photometry and showed the two values are consistent in a broad range, with  typical scatter of around 0.2 dex.
	
	The spatially resolved photometry is retrieved from the multi-band, multi-aperture photometry catalogue of CANDELS (Liu et al., in prep.), which measures the annulus-averaged radial surface brightness profiles in \emph{HST}/WFC3 bands and \emph{HST}/ACS bands. In brief, the aperture photometry is conducted after oversampling galaxy images in a certain band using \textsc{iraf} and  subtracting the sky background. For each galaxy, we always use elliptical annuli to run the aperture photometry with fixed shape parameters, including ellipticity, the position angle and the center position, which are determined from fitting WFC3/F160W images with \textsc{galfit}. Annulus radii along the ellipse major axis start with 0.04\arcsec, expanding outward in logarithmic steps of 0.1 dex, and end when the surface brightness is lower than 26 mag/arcsec$^2$. Therefore at 2 effective radii ($\sim$1.2\arcsec) of a typical galaxy in our sample, the separation between two consecutive apertures is around 5 times the drizzeld WFC3 pixel size (0.06\arcsec), allowing the radial flux profile to be determined by averaging a few tens of pixels. This procedure substantially reduces photometry noise at galaxy outskirts.
	
	To maximize the sample size and obtain accurate rest-frame colours at the same time, we use the photometry in two ACS bands (F606W and F814W) and two WFC3 bands (F125W and F160W) simultaneously, because they are the only four bands that are well covered both in UDS and GOODS-South. In order to avoid spurious colour gradients caused by the PSF (point spread function) mismatches among different bands, all images are convolved to match the resolution of WFC3/F160W (H band), which has a PSF FWHM of 0.18\arcsec. Rest-frame $U-V$ and $V-I$ colours from $z=0.4$ to $z=1.4$ are then derived from the four observed-frame bands using the empirical fitting relations derived Appendix \ref{append:A}. 	
	\begin{figure*}
		\centering
		\includegraphics[width=4.8in]{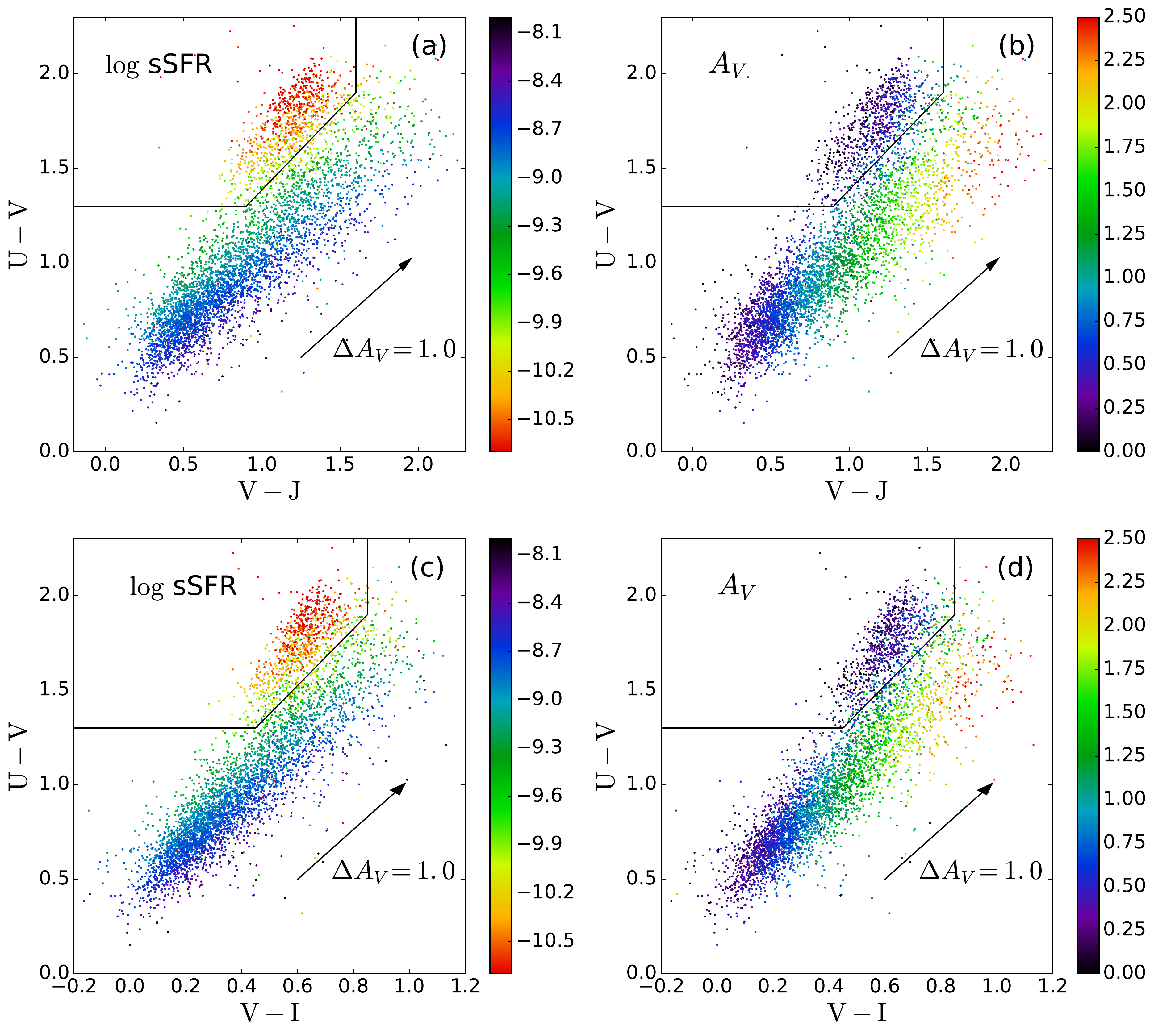}
		\caption{Two-colour diagrams for CANDELS/GOODS-South \& UDS galaxies at $z=0.4$--1.4. Rest-frame colours are derived from the \textsc{eazy} code \citep{Brammer2008}. \emph{Top}: \textit{UVJ} diagrams colour-coded by specific star-formation rate (sSFR) on the left and $A_V$ on the right. \emph{Bottom}: \textit{UVI} diagrams for the same galaxies colour-coded in the same way. The black lines serve as the boundary to separate quiescent galaxies from star-forming galaxies. The distribution of galaxies with different sSFR and $A_V$ values on the \textit{UVI} plane follows almost the same pattern as \textit{UVJ}.}
		\label{fig:UVI}
	\end{figure*}	
	A total of 1328 SFMS galaxies from the GOODS-South and UDS fields that satisfy the following criteria are selected. Sample sizes after each cut are listed in Table  \ref{tab:sample}.
	\begin{enumerate}
		\item WFC3/F160W band magnitude brighter than 24.5 mag. 
		\item \textsc{SExtractor} \citep{Bertin1996} parameter \textit{CLASS\_STAR} smaller than 0.9 to remove stars. 
		\item Both \textsc{galfit} and \textsc{SExtractor} flags equal to zero. This ensures that the WFC3/F160W images are not contaminated by star spikes, haloes, image edges, or artifacts and that the \textsc{galfit} fit is good (refer to \citealt{VanderWel2012} for details).
	
		\item Galaxy stellar mass $M_\star$ within the range of $M_\star=10^{9}$--$10^{11} \mathrm{M}_{\sun}$, and $z=0.4$--$1.4$. For galaxies beyond $z=1.4$, the rest-frame $V-I$ colour index cannot be obtained accurately due to a lack of photometry information redder than the WFC3/F160W band (see Appendix \ref{append:A}).
		
		\item Aperture photometry available in the four bands, ACS/F606W, ACS/F814W, WFC3/F125W, and WFC3/F160W.  
			
		\item Green-valley galaxies and quiescent galaxies are removed according to the positions on the sSFR-$M_{\star}$ plane. Following Fang et al.~(2017, submitted),  linear fittings of the sSFR-$M_{\star}$ relation are conducted in three redshift ranges, $0.2<z<0.5$, $0.5<z<1.0$ and $1.0<z<1.5$, using a sample of galaxies selected  by criteria 1, 2, 3 itemized above and $9.0<\log M_{\star}/\mathrm{M}_{\sun}<11.0$. The residual relative to these relations, $\Delta {\mathrm{sSFR}}$, must be above $-0.45$ dex in order to select only SFMS galaxies.
		
		\item Axis ratio $b/a>0.5$, where $a$ and $b$ are the effective radii along the galaxy projected semi-major and semi-minor axes, as obtained from the WFC3/F160W \textsc{galfit} fitting \citep{VanderWel2012}. This makes appropriate correction of PSF-smearing effect in the following sections possible, and also reduces reddening uncertainties by selecting out objects with high inclination (attenuation). We are aware that the axis ratio cut can seriously bias the sample towards intrinsically prolate or oblate objects, if our sample contains comparable amounts of these two geometry types and their ratio quickly varies with mass and redshift. However, as shown by \citet{VanderWel2014}, the prolate objects are only common (>20 per cent) in the mass-redshift range of $9.0<\log M_{\star}/\mathrm{M}_{\sun}<9.5$, $1.0<z<1.4$ (named as ``prolate range'' hereafter for convenience). In the other ranges where our selected galaxies lie, the prolate population is <20 per cent and the predicted variation of prolate galaxy fraction with mass and redshift is only around ten per cent. Therefore the considered selection bias should be modest in the ``prolate range'' and negligible for the other ranges. When the $b/a$ cut is made to the ``prolate range'' sample,  48.5 per cent of the galaxies are selected, a few per cents smaller than the percentage (53.4 per cent) selected to the whole sample, which can be explained by that slightly more prolate objects with small axis ratio are being weeded out compared with the whole sample. We stress that accurate modelling of 3-D geometry parameters is helpful to fully quantify the effect in future works.
		
		\item Effective radius along the semi-major axis, \textit{SMA}, in the WFC3/F160W band is larger than the typical F160W PSF size 0.18\arcsec (3 pixels), to make sure that the colour gradients are resolved. 
		
	\end{enumerate}	
	The final sample distribution on the F160W size-mass  plane is plotted as the red points in Fig.~\ref{fig:sample}. Galaxies satisfying criteria 1--4 and 6--8 but having angular sizes smaller than required by criterion 5 are plotted as gray points.  The angular size of \textit{SMA} spans from 0.30\arcsec to 0.85\arcsec, which is several times the PSF FWHM size, and the corresponding \textit{SMA}s are 2.5--5.5 kpc.  As seen from the plot, the angular size cut deletes more low-mass galaxies, but almost all of the discarded objects lie beyond the $1\sigma$ range of the size-mass relation. This lack of sample completeness in our work cannot be avoided given the limited available resolution.	
	\section{The UVI colour diagram} 
	\label{sec:UVI}
	
	Two-colour diagrams are frequently used in analyzing the character of stellar population ages and dust extinctions. A common example is rest-frame $U-V$ versus $V-J$ (e.g., \citealt{Labbe2005, Wuyts2007, Williams2009, Patel2011b}). However, our reddest data being limited to 1.6 \micron\ (WFC/F160W), rest-frame $J$ band photometry becomes increasingly uncertain beyond $z\sim0.4$. We therefore explore the possibility of replacing $J$  with a bluer rest-frame band to construct an alternative two-colour diagram.

	Fig.~\ref{fig:UVI} shows the distribution of a selection of galaxies on the $U-V$ versus $V-I$ plane (hereafter \textit{UVI}), as well as in \textit{UVJ} for comparison. All CANDELS GOODS-South and UDS galaxies with $0.4<z<1.4$, $9.0<\log M_{\star}/\mathrm{M}_{\sun}<11.0$, and filtered by criteria 1, 2, 3 in Section \ref{sec:data} are shown. Each point in the plot represents the integrated colours of an individual galaxy. The colour codings show the distribution patterns of sSFR and $A_V$ in UVI and UVJ, the latter initially presented by Fang et al.~(2017, submitted) and reproduced here for reference. It is seen that $UVI$ reproduces all the main features of \textit{UVJ}, including the quenched region and the same distinctive patterns of sSFR and $A_V$ stripes. 
	
	The agreement between \textit{UVI} and \textit{UVJ} is consistent with the smooth slope and small cosmic scatter of galaxy SEDs in the $V$-$I$-$J$ range. A drawback of the substitution is that \textit{UVI} has smaller dynamic range and thus places somewhat poorer constraints on sSFR and $A_V$ than \textit{UVJ}. The separation between the quenched population and the star-forming region, as well as the width of SFMS ridge-line, is therefore smaller in the \textit{UVI} diagram than in \textit{UVJ}. 
	However we will estimate the associated uncertainty in Section \ref{sec:UVIcali} below and validate that \textit{UVI} is a satisfactory 
	substitute for \textit{UVJ}.
	
	Fig.~\ref{fig:UVI} shows boundary lines that separate quenched galaxies from the SFMS population. The boundary lines in \textit{UVJ} are the same as those adopted for $1.0<z<2.0$ by \citet{Williams2009} with the redshift dependence neglected in this work for simplicity. For \textit{UVI}, the boundary lines are determined by visually confirming that objects on the \textit{UVJ} boundaries also lie on the boundary lines in \textit{UVI}.

	\begin{figure*}
		\centering
		\includegraphics[width=7in]{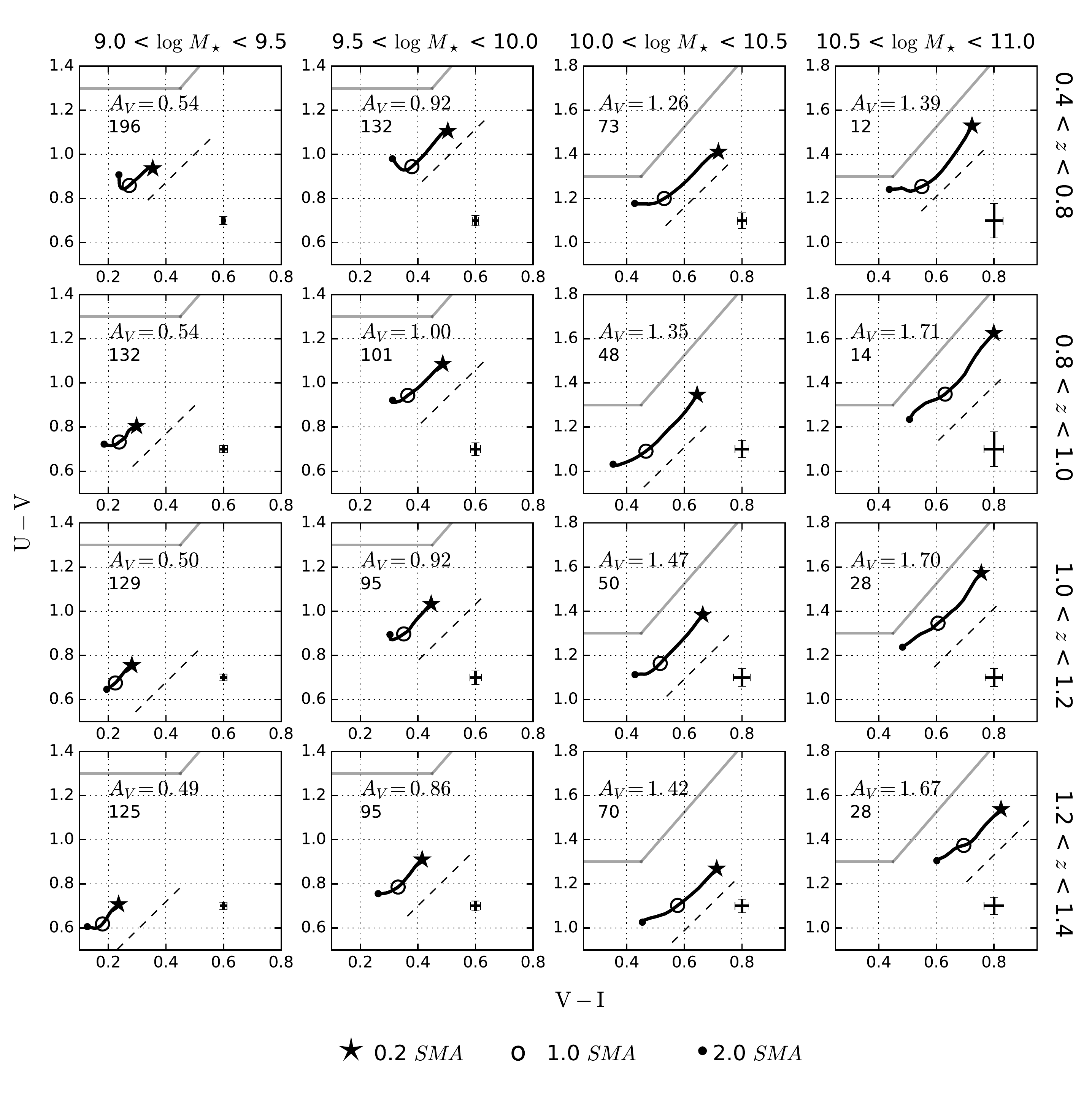}
		\caption{UVI colour gradients for star-forming galaxies in mass-redshift bins, as  indicated at the top and right margins. The radial colour profiles are scaled by effective radius before stacking, without correction for PSF smearing. For each panel, the stacked radial trajectories in $U-V$ and $V-I$ are plotted as the black curves. Dashed lines denote the Calzetti reddening vector. Mean \emph{integrated} $A_V$ and the number of sample galaxies are indicated at upper left in each panel, and the typical standard deviation of the mean of \emph{stacked} gradients  is shown at lower right. The gradient trajectories follow the reddening vector, which indicates that most of the gradients are caused by dust variations and that sSFR is nearly constant with radius.}
		\label{fig:UVIgradient_re}
		
	\end{figure*}

	\section{UVI colour gradients} \label{sec:UVIgradient}
	
	\subsection{Sample binning, PSF smearing effect, and the stacking methods}
	\label{sec:UVIgradientPSF}
	Since galaxies evolve with mass and redshift, we divide the sample into four stellar mass bins, $10^{9.0}$--$10^{9.5}\mathrm{M}_{\sun}$, $10^{9.5}$--$10^{10.0}\mathrm{M}_{\sun}$, $10^{10.0}$--$10^{10.5}\mathrm{M}_{\sun}$, and $10^{10.5}$--$10^{11.0}\mathrm{M}_{\sun}$, and four redshift bins, 0.4--0.8, 0.8--1.0, 1.0--1.2, and 1.2--1.4 and stack galaxy colour profiles in the same mass-redshift bin to reduce photometry uncertainty and cosmic scatter. However, there are two challenges to be considered. First, the light smearing effect caused by PSF on colour gradients must be evaluated carefully, especially within 0.18\arcsec (3 times the drizzled WFC3 pixel size). Second, even in the same mass-redshift bin, galaxies can have non-negligible physical size scatter, which may make the derived average colour profiles seriously dependent on the choice of the corresponding positions to overlay the profiles. Our strategy to deal with these is to generate and compare multiple versions of colour gradients using different sample stacking methods (stacking galaxy colour profiles on a fixed size scale in arcsec, versus stacking colour profiles after individual galaxy profiles are normalised by their effective radii), and different treatments of PSF smearing effect (no treatment versus correction based on a variety of  prior models). We also do not calculate colour gradients inside 0.2 effective \textit{SMA} (around 0.7 kpc, or $\sim$ 0.06\arcsec), where PSF smearing is overwhelmingly important.
	
	Since the photometry in all bands has been convolved to match the WFC3/F160W resolution (\citealt{Guo2013}), any influence caused by different PSF sizes of \emph{HST} filters has been eliminated. All galaxies receive the same statistical weight in a stack. The maximum radial extent to measure colour gradients is set to 2 times the \textit{SMA}, where reliable photometry is still available.

	\subsection{Colour gradients uncorrected for beam smearing and stacked by radius scaled to \textit{SMA}}
	\label{sec:gradient_re}
	
	This subsection shows raw stacked colour gradients without any beam-smearing correction. First, we scale the radial colour profiles of each individual galaxy by its SMA (effective radius along the galaxy major axis), by interpolating between the apertures as described in Section \ref{sec:data}. Then these scaled colour profiles are stacked together to produce the average profiles as shown in	Fig.~\ref{fig:UVIgradient_re}.

	The grid of panels in
	Fig.~\ref{fig:UVIgradient_re} shows the shapes of the stacked colour trajectories in \textit{UVI}-space as
	a function of mass and redshift.  Each panel is a \textit{UVI} diagram, and the
	black lines are the paths taken by the colours in going from the innermost radius
	(upper right) to the outermost radius (lower left). The three symbols indicate local colours at 0.2 \textit{SMA} (star),  1 \textit{SMA} (open circle), and 2 \textit{SMA} (dot). The direction of dust reddening by the Calzetti law is shown
	as the dashed lines. 
	
	The main feature of this figure is that gradient trajectories approximately
	follow the reddening vector, which indicates that most of the gradients
	are caused by dust variations and that sSFR is nearly constant with radius. This conclusion is unchanged under beam-smearing correction and stacking by radius rather than scaled radius, as shown in the next section.
	\begin{figure*}
		\centering
		\includegraphics[width=7in]{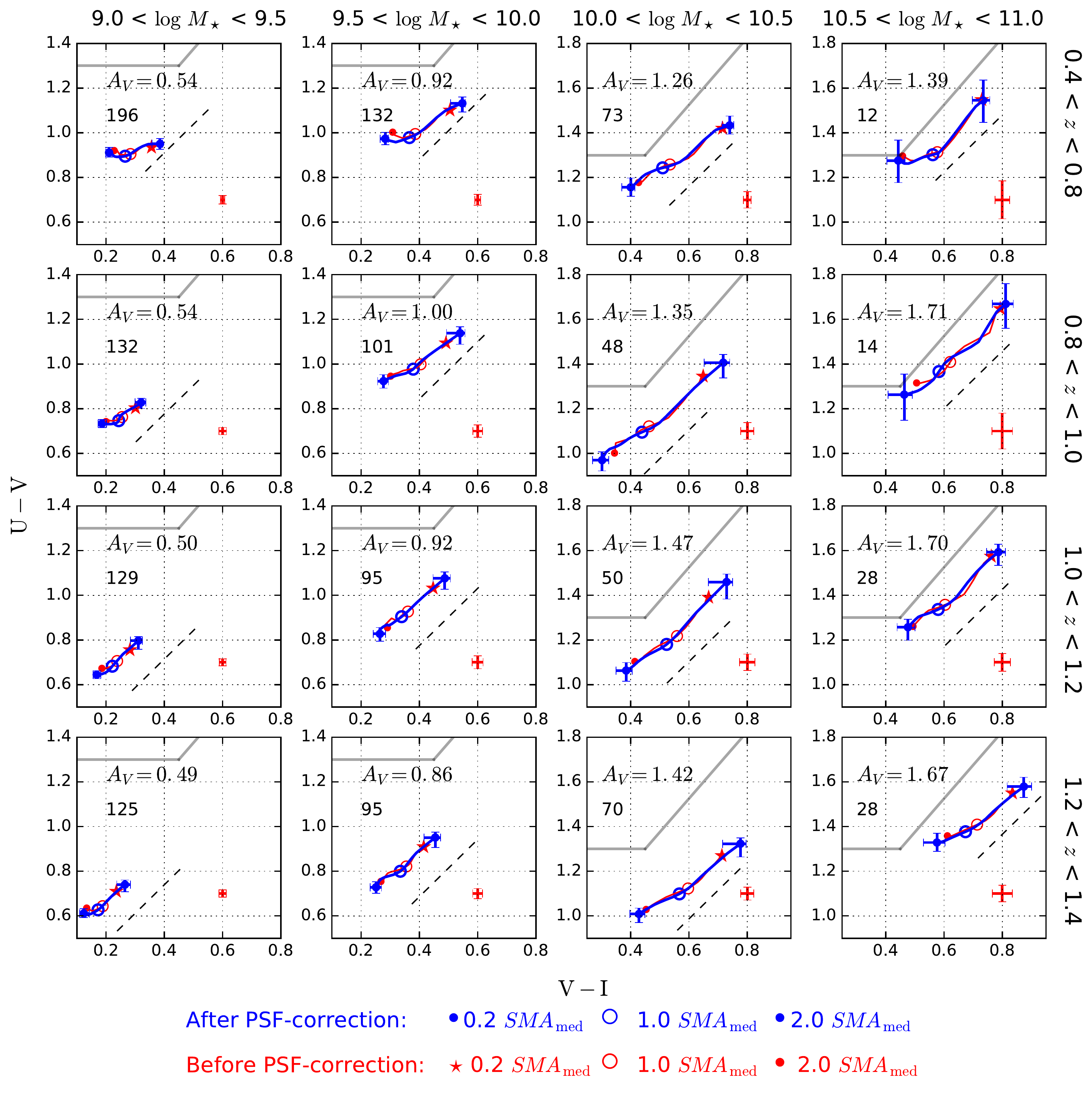}
		\caption{ This figure repeats the \textit{UVI} colour gradients shown in Fig.~\ref{fig:UVIgradient_re} but now
			with colour profiles stacked based on the angular distance (in arcsec) to galaxy centers. 
			The red curves are the raw gradients, while the blue curves
			are the colour gradients after PSF-correction, as explained in Fig.~\ref{fig:PSFcorrection}.
			Symbols, as annotated at the figure bottom, are the radial distances
			along the tracks in units of $\mathit{SMA}_{\,\mathrm{med}}$, the \emph{median} angular size of galaxy \textit{SMA}s in each mass-redshift bin.
			The red error bars are the typical standard deviation of the stacked gradients. The blue error bars at the ends of the curves
			have had PSF correction uncertainties added in quadrature, defined as the maximum deviation of the gray curves
			from the blue curves in Fig.~\ref{fig:PSFcorrection}.
		}
		\label{fig:UVIgradient}
	\end{figure*}

	\subsection{Colour gradients stacked based on angular distance and beam-smearing correction}
	\label{sec:psfcorr}

	Next, we turn to another stacking method of colour profiles and then evaluate and correct the beam-smearing effects caused by the PSF. 
		
	To start with, galaxy colour profiles are stacked together based on the {\em{angular distance in arcsec}} to their centers, instead of the nomalisation used in Section \ref{sec:gradient_re}. The resultant colour gradients are shown as red curves in Fig.~\ref{fig:UVIgradient}. This stacking method ensures that the stacked profiles have a well-defined angular resolution at each radius, which is necessary to conduct the following PSF correction. We are aware that this method holds out risk that different galaxy parts are stacked together for galaxies with different angular sizes.  However, fortunately when comparing these colour profiles (red curves in Fig.~\ref{fig:UVIgradient}) with those stacked by scaled radii (Fig.~\ref{fig:UVIgradient_re}), no significant difference is found, showing that this issue does not alter our principal conclusions.
	
	Regarding the PSF smearing effect, the observed shapes of galaxy colour profiles can be different from the intrinsic ones even though all the corresponding band images have been PSF-matched. Qualitatively, for a galaxy with a light distribution that is more centrally peaked in one band than another, more light will be smeared out from the centre of that band when convolved by the same PSF function. Because the presence of colour gradients is indicating that the light in one band is indeed more centrally concentrated than in another band, the PSF smearing effect across different bands will cause the observed colour gradients to be smaller than the intrinsic ones. 
	
	We model and correct this effect following \citet{Szomoru2010}. This method requires a prior model of the intrinsic light distribution, and the final light profiles are generated by adding the residual from fitting the observed light profiles with the PSF-convolved model on to the prior profiles. Although a prior assumption is always used, it has been shown that the final result is insensitive to the choice of prior, and the method also avoids false features caused by noise, as can occur in the direct deconvolution method. 
	
	To implement this, the redder-band image of each filter pair is modeled as a single S\'{e}rsic component. The median values of the S\'{e}rsic indexes, effective radii, and ellipticities of the galaxies in the stack, as retrieved from the CANDELS WFC3/F160W morphology catalogue \citep{VanderWel2012}, are used to construct this model.  The bluer-band image is modeled by adding various prior colour gradient components on to the red image.  These prior gradients are assumed to be proportional to $\log r$ except for the central regions, with slope values obtained from fitting the observed colour gradients between 0.6 \textit{SMA} and 1.5 \textit{SMA}, where beam-smearing effects are small. Flat truncation or linear interpolation of the prior colour gradients is adopted within 0.05/0.1/0.2/0.3/0.4/0.5/0.6 \textit{SMA} to avoid unphysical logarithmic divergence at the centre and to cover all physically probable conditions. The \citet{Szomoru2010} procedure is then applied to all the prior models, and the multiple colour gradients are generated based on different priors. For a certain mass-redshift bin, the median values of these generated curves  are taken as the our PSF-corrected colour profiles, while the maximum deviations from the medians are retained as PSF-correction uncertainties.
	
\begin{figure*}
			\centering
			\includegraphics[width=4in]{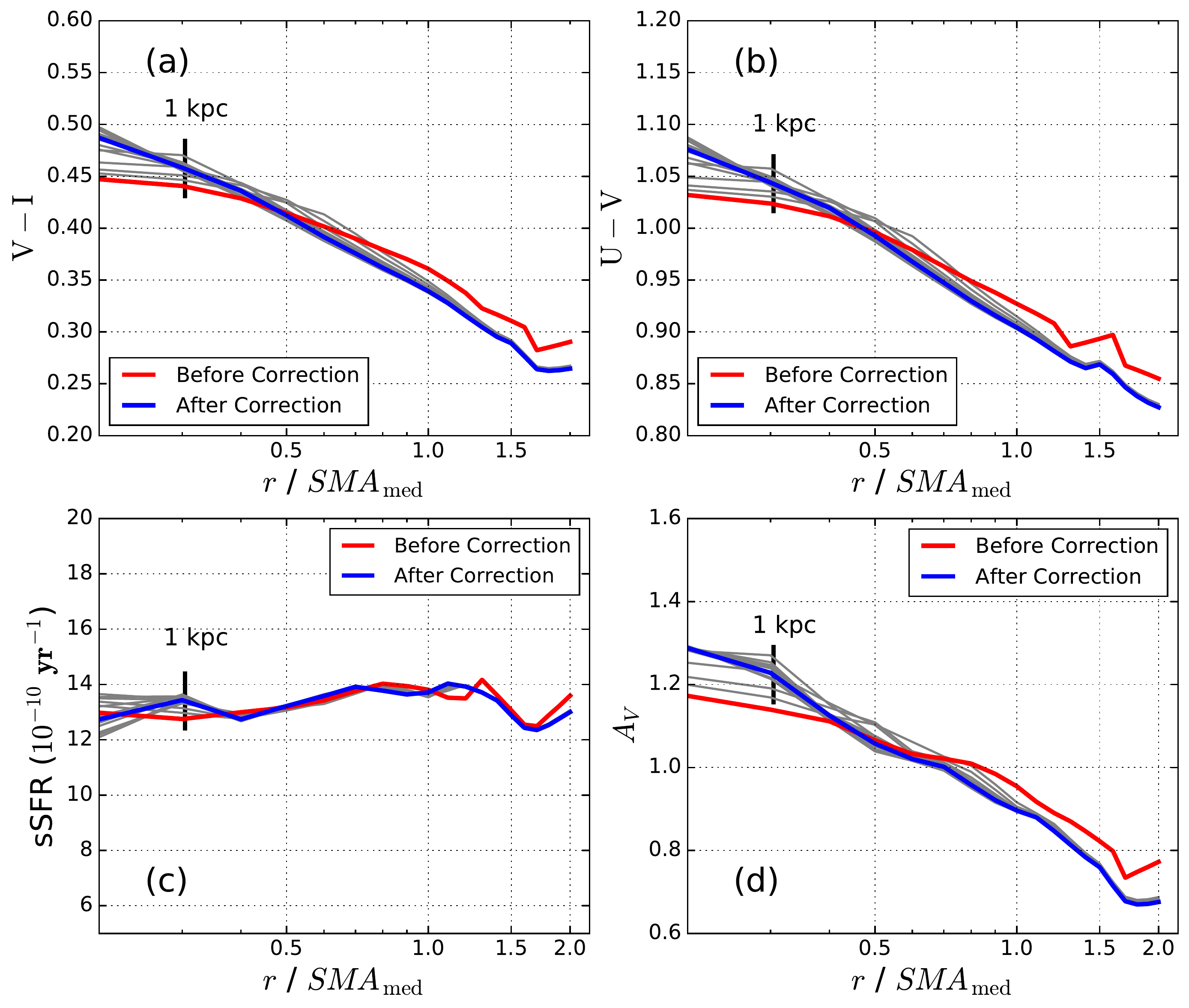}
			\caption{An example for the correction of PSF smearing effect on to the stacked radial gradients, for the mass-redshift bin $9.5<\log M_\star/\mathrm{M}_{\sun}<10.0$ and $1.0<z<1.2$ (see the main text for details).  Galaxies are stacked according to the angular galactocentric distance in arcsec, but the $x$-axis is plotted as a fraction of the median angular effective radius of the stacked galaxies ($\mathit{SMA}_{\,\mathrm{med}}$). \emph{Top}: Original measured $U-V$ and $V-I$ gradients are shown as red curves, and gray are the corrected curves assuming various prior models. Blue are the \emph{median values} of the gray curves, which is adopted throughout the paper as `PSF corrected colour gradients' (See Section \ref{sec:psfcorr}). \emph{Bottom}: The corresponding sSFR and $A_V$ radial profiles derived from colour gradients, using the \textit{UVI} calibration method as will be shown in Section \ref{sec:UVIcali}. It is seen that the original effects of beam-smearing and their corrections are both small.} 
			\label{fig:PSFcorrection}
	\end{figure*}

	For the purpose of demonstration, an example from the stack with $9.5<\log M_\star/\mathrm{M}_{\sun}<10.0$ and $1.0<z<1.2$ is shown in Fig.~\ref{fig:PSFcorrection}.    As shown in the two upper panels of Fig.~\ref{fig:PSFcorrection}, the corrected profiles are systematically redder in the centre and mildly bluer in the outskirts. 
	
	The resultant PSF-corrected colour profiles for all redshift and mass bins are plotted as the blue curves in Fig.~\ref{fig:UVIgradient}, with error bars indicating the combined uncertainty of the PSF-correction and the standard deviation of the stacked profile. As expected, the PSF-correction stretches the length of colour gradients on the \textit{UVI} diagram by making galaxy colours redder in the centre and bluer in the outskirts. However, the overall effect of PSF-correction is not large.
	
	\subsection{Basic findings on stacked colour gradients}
	
	By comparing the overplotted red and blue curves in Fig.~\ref{fig:UVIgradient} with Fig.~\ref{fig:UVIgradient_re}, we are able to capture certain basic features in common. These features, as we have shown in Section \ref{sec:gradient_re} and Section \ref{sec:psfcorr}, are robust against different treatments of PSF-correction and galaxy-radius stacking. They are the foundation for drawing further scientific conclusions in the rest of this article. 
	
	First, the most striking aspect of the figures is that the gradient trajectories
	are all roughly parallel to the reddening vector (except for small upturns near 2 \textit{SMA}, which are discussed below).  
	
	Second, all points lie well within the star-forming region on the \textit{UVI} plane, indicating that on average
	the centres of galaxies are not observed to have quenched at our raw resolution 
	of $0.18\arcsec$, which is typically 1.5 kpc at $z\sim 1$.  
	This can be extended down to 0.2 \textit{SMA} ($\sim 0.6$ kpc) using PSF-correction. Fig.~\ref{fig:PSFcorrection} demonstrated that PSF-corrections are small even at this radius, indicating that this region is well resolved.  Nearby bulges are roughly of this size ($R_e\sim 0.7$ kpc, \citealt{Gadotti2009, Delgado2015, Delgado2016}). If distant bulges have the same radial extent as local bulges and if they are fully quenched at the epochs we are observing them, their centres should lie inside the quenched region of the \textit{UVI} diagram,  but this is not seen, either for the raw gradients or the corrected gradients.  Instead, central PSF-corrections tend to extend a bit further inward the dust trends seen at larger radii.
	
	Third, we find that the magnitude of the colour gradients increases significantly with mass at low $M_{\star}$ ranges, especially from $10^{9.5}$--$10^{10.0}\mathrm{M}_{\sun}$ to $10^{10.0}$--$10^{10.5}\mathrm{M}_{\sun}$.  But from  $10^{10.0}$--$10^{10.5}\mathrm{M}_{\sun}$ to $10^{10.5}$--$10^{11.0}\mathrm{M}_{\sun}$, the strength of colour gradients tends to saturate.  Considering that $10^{10} \mathrm{M}_{\sun}$ is observed to be near the pivot mass of disk settling at intermediate and high redshifts \citep{Kassin2012, Simons2016}, the strength of colour gradients, as an indicator of the radial variation of underlining stellar population or dust/ISM distribution, may be sensitive to and correlated with disk formation. Elongation along the Calzetti vector increases with both mass and global $A_V$ at fixed redshift, but the increases with time at fixed mass are small.  This is in agreement with previous work showing
	that dust content increases strongly
	with mass (e.g., \citealt{Reddy2006, Reddy2010}; \citealt{Pannella2009}).
	To first order, $A_V$ is well parametrized by mass alone in our data.
	Nevertheless, we will show in Fig.~\ref{fig:avprofile} that massive 
	galaxies arriving at a given mass at late times (e.g., $10^{10}$--$10^{11} \mathrm{M}_{\sun}$
	and $z  = 0.4$--0.8) have slightly \emph{less} $A_V$ for their mass, perhaps
	reflecting a loss of ISM as star formation shuts
	down.
	
	The fourth noteworthy feature of Fig.~\ref{fig:UVIgradient_re} and Fig.~\ref{fig:UVIgradient} is the smoothness of the gradient
	trajectories.  The number of galaxies averaged in
	each bin is evidently adequate to smooth out the measurement noise in the
	individual profiles, along with the effects of any local structures like clumps (e.g., \citealt{Elmegreen2005, Wuyts2012, Guo2012, Guo2015}).  
	Assuming that the underlying calibration
	of \textit{UVI} to dust and star formation is also smooth, this implies that the
	resulting radial curves of sSFR and $A_V$ will also be smooth.
	This proves to be the case, as will be shown in Section \ref{sec:radialprofile}.

	The fifth feature of Fig.~\ref{fig:UVIgradient_re} and Fig.~\ref{fig:UVIgradient} is the slight upturns in the paths that are visible in the extreme outer parts of galaxies near 2 \textit{SMA}.  Closer inspection shows 
	these upturns to be about equally prominent at all masses but to be stronger at late times (the lowest mass bin at lowest redshift is particularly strongly affected).  The origin of these upturns is unknown. Probable interpretations can be that galaxies are surrounded by old stars that
	were scattered from central parts due to dynamical migration \citep{Roskar2008} or accreted via minor mergers. Similar age or colour gradient turnover have
	been noted for local galaxies (e.g. \citealt{Bakos2008, Radburn-Smith2012, D'Souza2014, Zheng2015}).  Also \citet{Wuyts2012} finds disks with older stellar ages in the outskirts at $z\sim 1$.
	The direction of the upturn indicates older stars, and this will be evident as a 
	downturn in sSFR at the outer radii of our galaxies at late times. The existence of these upturns and their uncertain interpretation is the reason why we have truncated our analysis at 2 \textit{SMA}. Further consideration is beyond the scope of this paper.

	\section{$A_V$ and sSFR calibrations}
	\label{sec:UVIcali}
	\begin{figure}
		\centering
		\includegraphics[width=3in]{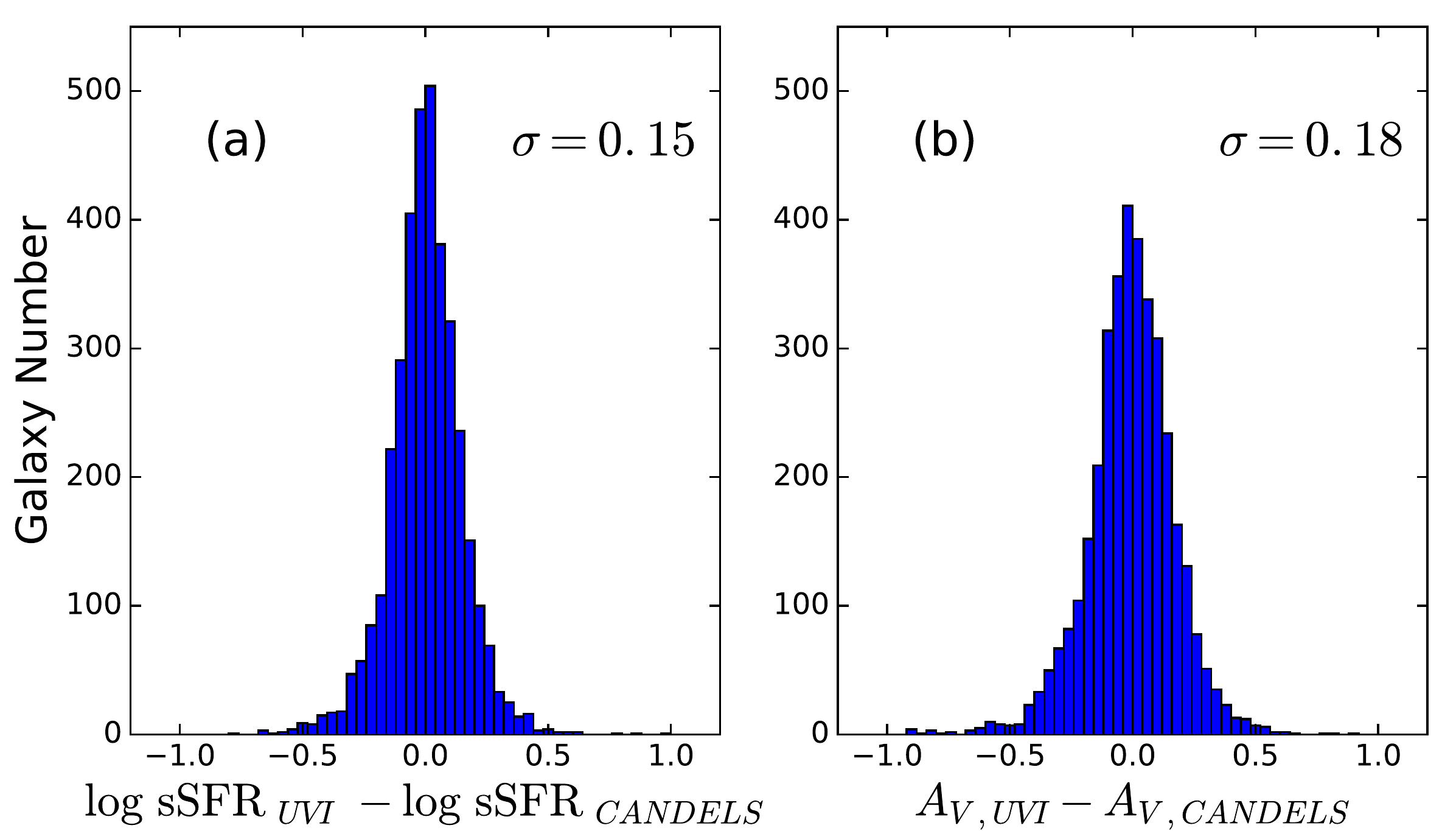}
		\caption{ A comparison between the original sSFR and $A_V$ values obtained from the CANDELS catalogue ($\mathrm{sSFR}_{\mathit{CANDELS}}$ and $A_{\mathit{V, CANDELS}}$), and the sSFR and $A_V$ derived from galaxy positions on the \textit{UVI} diagram ($\mathrm{sSFR}_{\mathit{UVI}}$ and $A_{\mathit{V, UVI}}$). Histograms of the residuals are shown, with standard errors of the distribution $\sigma$ annotated at the upper right.}
		\label{fig:cellscatter}
	\end{figure}
	
	We turn now to the task of converting raw \textit{UVI} colours into values of $A_V$ and sSFR.
	As shown in Fig.~\ref{fig:UVI}, galaxies on the \textit{UVI} diagram exhibit clear sSFR and $A_{V}$ patterns with small scatter. These ordered patterns make it possible to deduce sSFR and $A_{V}$ values from the $U-V$ and $V-I$ of a whole galaxy or of any part of a galaxy.  
	
	We start by creating a map of sSFR and $A_V$ in the \textit{UVI} plane. We do this by plotting integrated \textit{UVI} colours as in Fig.~\ref{fig:UVI}c and Fig.~\ref{fig:UVI}d, and dividing the resulting distribution into multiple $0.05\times0.10$ mag rectangles.  In each rectangle we calculate the median CANDELS catalogue values of sSFR and $A_V$ and assign them to the centre of the rectangle.  In this way, maps of sSFR map and $A_{V}$ are generated based on the integrated properties of the galaxies. Then, given any position on the \textit{UVI} plane, the corresponding sSFR and $A_{V}$ values are obtained by linearly interpolating among the nearby rectangle centres. 
	
	To assess the results, we derive a set of sSFR and $A_V$ values for all galaxies using integrated \textit{UVI} colours as in Fig.~\ref{fig:UVI} ($\mathrm{sSFR}_{\mathit{UVI}}$ and $A_{\mathit{V, UVI}}$) and compare them to the original values in the CANDELS catalogue ($\mathrm{sSFR}_{\mathit{CANDELS}}$ and $A_{\mathit{V, CANDELS}}$). Histograms of the difference between the two set of parameters are shown in Fig.~\ref{fig:cellscatter}. As annotated in the figure, sSFR and $A_V$ can be constrained using \textit{UVI} with an rms accuracy of 0.15 dex and 0.18 mag respectively.
	
	\section{Radial profiles}
	\label{sec:radialprofile}
	\subsection{Radial $A_V$ profiles}
	\begin{figure*}
		\centering
		\includegraphics[width=5in]{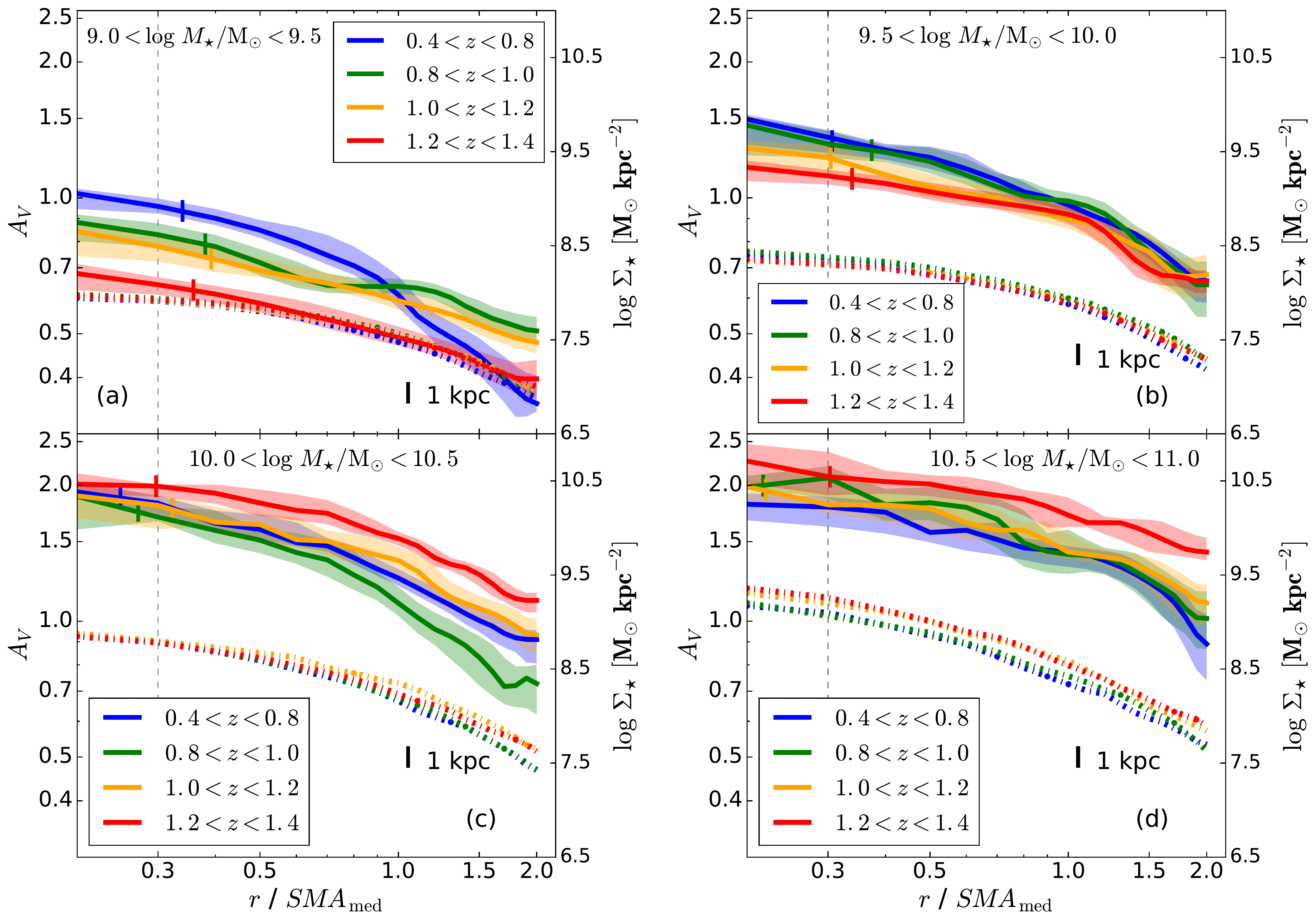}
		\caption{$A_V$ profiles of face-on ($b/a>0.5$) galaxies derived from the stacked \textit{UVI} colour gradients (blue curves in Fig.~\ref{fig:UVIgradient}). Solid curves with different colours demonstrate different redshifts, and mass ranges are annotated at the top of each panel. Coloured shadows show the profile errors including PSF-correction uncertainties (see text). Radial distances in arc seconds are rescaled to \textit{SMA} units using the galaxy median angular \textit{SMA} size in each mass-redshift bin.  Dashed vertical lines lie at 0.3 \textit{SMA},  which approximate the typical WFC3/F160W PSF radius size ($\approx 0.18$\arcsec/2). The short coloured vertical lines show the median position of 1 kpc away from galaxy centres. Dashed curves with the same colours represent stellar mass density profiles derived from \textsc{fast}, assuming solar metallicity, Calzetti law and exponentially declining SFHs (note the separate mass density scales at the right axis). Normalization varies with mass and redshift, but the shape of $\log A_V$ vs. $\log r$ is quite constant over all galaxies.}
		\label{fig:avprofile}
	\end{figure*}
	
	Using the calibration method developed in Section \ref{sec:UVIcali}, the PSF-corrected \textit{UVI} colour profiles (blue curves in Fig.~\ref{fig:UVIgradient}) are converted to radial $A_V$ profiles, as shown in Fig.~\ref{fig:avprofile}. 
	Radial distances in arc seconds are rescaled to \textit{SMA} units using the \emph{median angular size} of the stacked galaxies in each mass-redshift bin. Coloured vertical line segments indicate the median physical position of 1 kpc. Shadowed regions account for the scatter caused by the stacked colour profile uncertainties (i.e., standard deviation of the mean) combined in quadrature with the PSF correction uncertainties shown in Fig.~\ref{fig:PSFcorrection}.  
	
	Fig.~\ref{fig:avprofile} bins galaxies by mass and colour-codes the $A_V$ profiles by redshift. A steady increase of the attenuation value, as well as the size of the radial gradient, with stellar mass at fixed redshift is observed. In contrast, redshift trends are much milder.  Note that the conventional analysis method we are using for $A_V$ and sSFR predicts dust attenuation values as high as $1.0$--$2.5$ mag in galaxy centres and up to 1.5 mag at the outskirts.  These high values, if real, mean that our optical view of SFMS galaxy structure is highly distorted by dust and that recovering accurate galaxy structural parameters depends on an accurate dust correction.  We return to this topic in Section \ref{sec:compareother}.
	
	\begin{figure}
		\centering
		\includegraphics[width=\columnwidth]{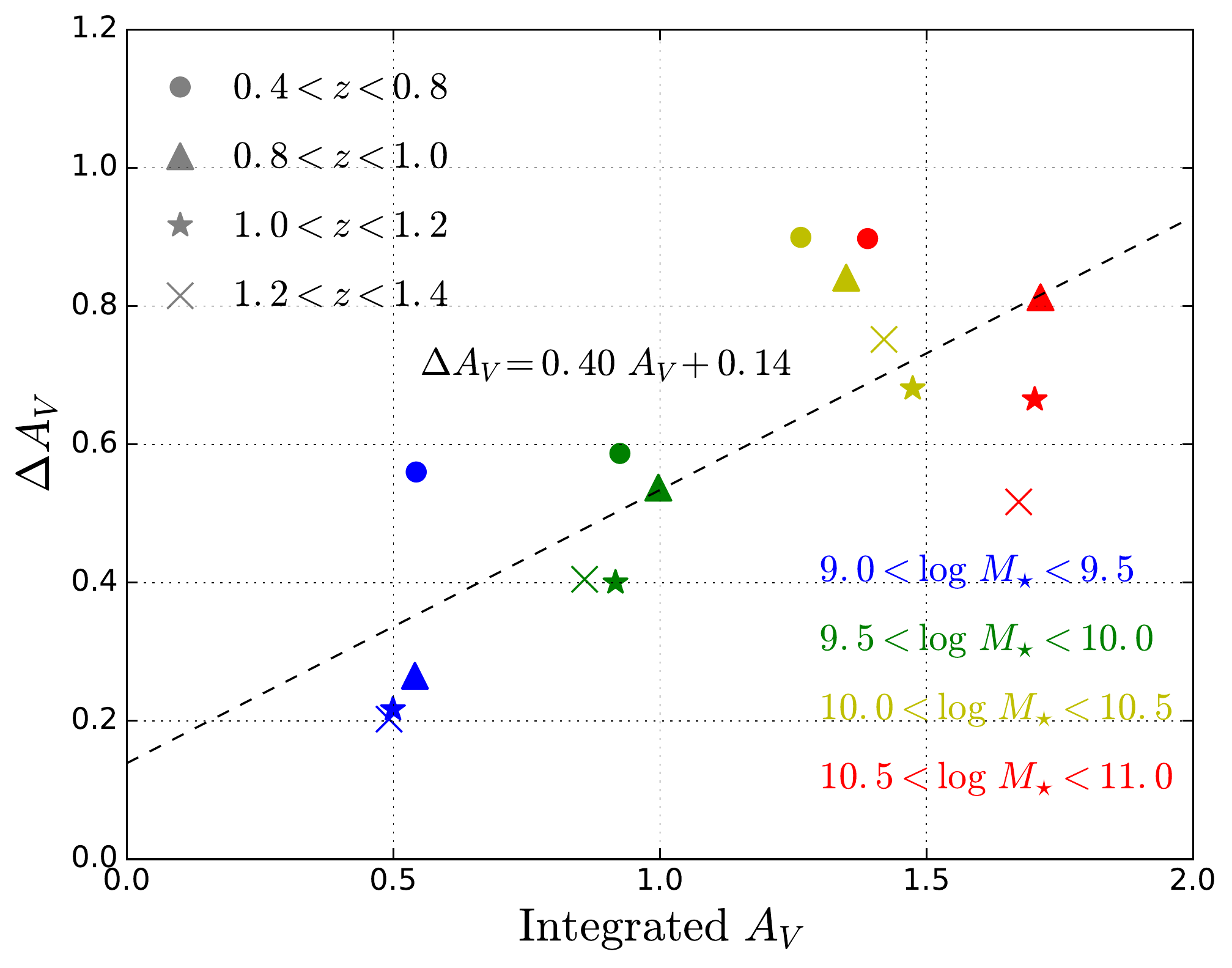}
		\caption{ Gradient length $\Delta A_V$, defined as the difference of $A_V$ values at 0.2 \textit{SMA} and 2 \textit{SMA}, versus galaxy integrated extinction values $A_V$ for different masses and redshifts.
			$\Delta A_V$ is roughly proportional to
			$A_V$, with the further trend that $\Delta A_V/A_V$ is growing moderately with time.  The
			proportionality indicates
			a roughly constant pattern in the radial shape of the absorption curve,
			as shown in Fig.~\ref{fig:avprofile}.
		}
		\label{fig:delav_av}
	\end{figure}

	\begin{figure*}
		\centering
		\includegraphics[width=5in]{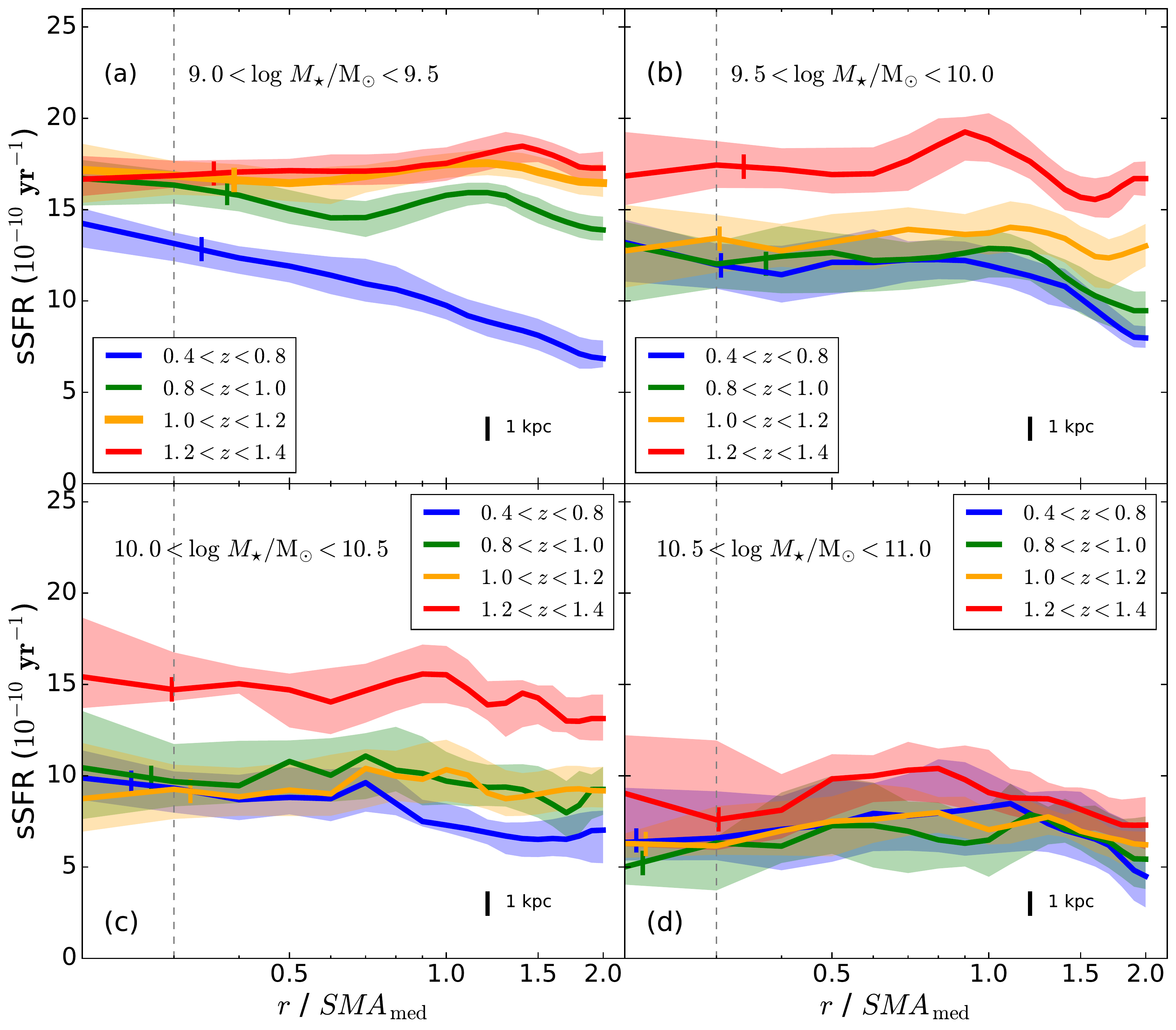}
		\caption{Specific star formation rate (sSFR) profiles derived from the same stacked colour gradients as mentioned in Fig.~\ref{fig:avprofile}. Curves with different colours show different redshifts, and mass ranges are annotated at the top of each panel. Coloured shadows stand for the standard uncertainty of the \emph{stacked} profiles. Radial distance is scaled by the galaxy median angular \textit{SMA} size of each mass-redshift bin. Dashed vertical lines lie at 0.3 \textit{SMA}, which approximate the typical WFC3/F160W PSF radius size ($\approx 0.18\arcsec/2$). The short \emph{coloured} vertical line segments demonstrate the median position of 1 kpc away from galaxy centres. Aside from the lowest mass bin at low redshift, the remainder of the profiles are basically flat with radius. A central dip of 20 per cent may exist at the highest mass.}
		\label{fig:ssfrprofile}
	\end{figure*}
	
	A measure of the gradient size, $\Delta A_V$, is plotted versus galaxy global $A_V$ in Fig.~\ref{fig:delav_av}.  $\Delta A_V$ is defined as the difference in $A_V$ between 0.2 \textit{SMA} and 2.0 \textit{SMA}.
	The points are colour-coded by mass and redshift, and it is seen that a straight line is tolerable fit with small Y intercept.
	This rough proportionality indicates a consistent pattern in $A_V$ vs. radius
	that is repeated from bin to bin.   Over the measured radii, $\Delta A_V$ averages about 0.4 $A_V$, and
	also increases slightly with time.  Aside from this latter point, the rough proportionality means that the development of dust inhomogeneity comes along with dust production in a concordant fashion.
	
	However, a major new finding in Fig.~\ref{fig:avprofile} is the ratio of $A_V$ to stellar mass surface density, $\Sigma_\star$. Median $\Sigma_\star$ profiles are shown as the dotted lines  in Fig.~\ref{fig:avprofile} based on stellar mass profiles obtained by Liu et al.\ (in prep.) from fitting \textsc{fast} \citep{Kriek2009} to the multi-band aperture photometry profiles. It is seen that stellar mass typically falls by a factor of 10--20 from centre to edge whereas $A_V$, which is strictly proportional to dust surface density under the foreground screen assumption, drops by a factor of 2. The dust to stellar mass ratio therefore increases outward by roughly one dex. This is perhaps unforeseen behavior and is discussed further in Section \ref{sec:avcritique}.

	\subsection{Radial sSFR profiles}
	Radial sSFR profiles accompanying the $A_V$ profiles in Fig.~\ref{fig:avprofile} are plotted in Fig.~\ref{fig:ssfrprofile}. The profiles of high-redshift galaxies are seen to lie above those of low-redshift galaxies, reflecting the overall drop in the level of the SFMS since $z\sim 2$ (e.g., \citealt{Noeske2007, Whitaker2012, Speagle2014}).
	More striking is the small variation in sSFR with radius.  With the exception of a single deviant curve at small masses and late times, sSFR profiles below $10^{10.5} \mathrm{M}_{\sun}$ are flat. This was foreshadowed by the \textit{UVI} gradient vectors in Fig.~\ref{fig:UVIgradient_re} and Fig.~\ref{fig:UVIgradient}, and is also consistent with the results of \citet{Liu2016}. These galaxies are evidently forming stars at virtually the same relative rate at all radii.  In the most massive bin above $10^{10.5} \mathrm{M}_{\sun}$, a central drop of around 20 per cent is seen, which may indicate the onset of central star-formation quenching associated with bulge formation in massive galaxies. This is discussed further in Section \ref{sec:ssfrcritique}.

	\section{Comparison with two recent works}
	\label{sec:compareother}
	\subsection{$A_V$ profiles}  
	\begin{figure*}
		\centering
		\includegraphics[width=3.7in]{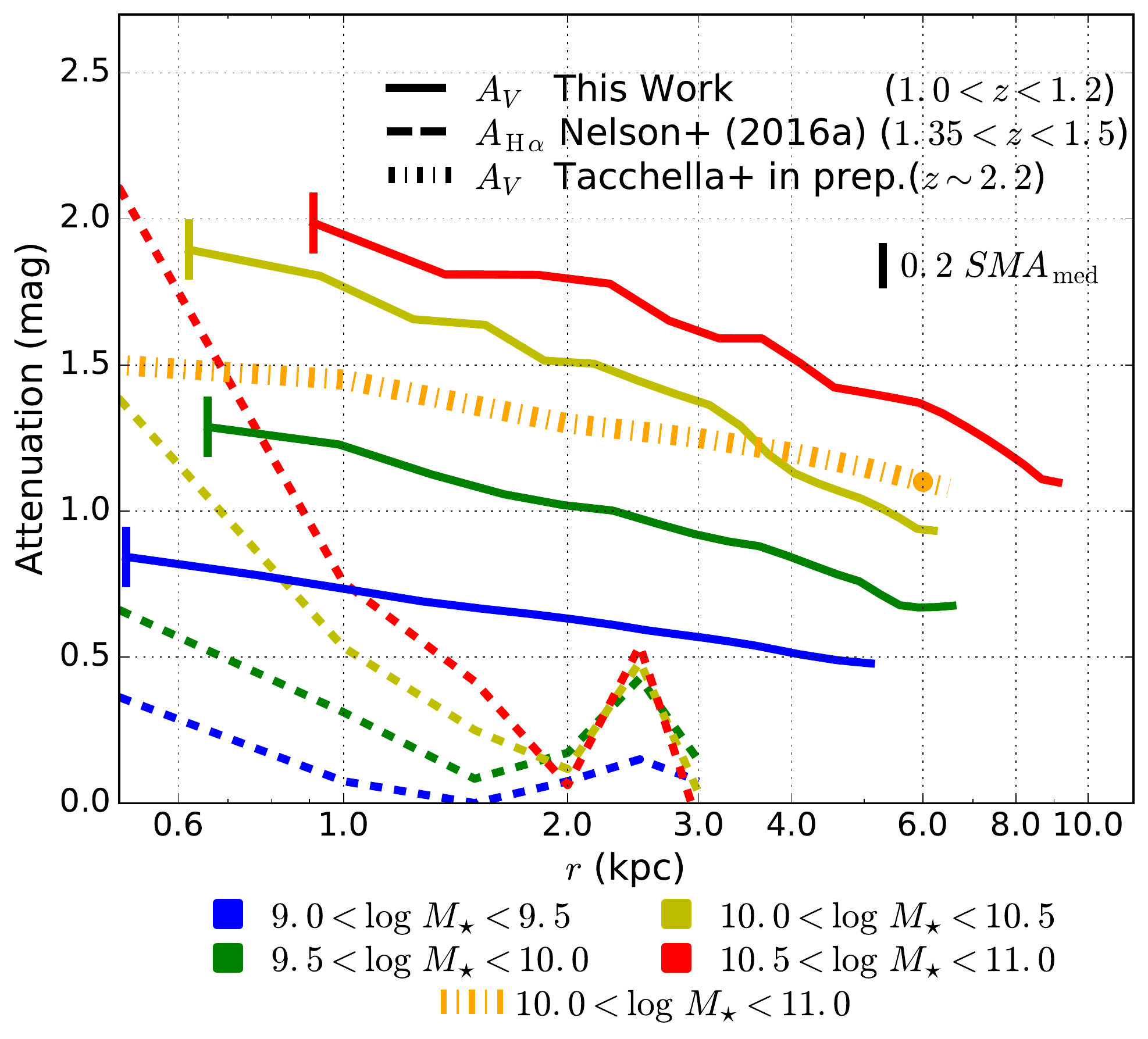}
		\caption{Dust extinction profiles from three different works, our \textit{UVI} colour gradient calibration, the \citet{Nelson2016} Balmer decrement measurement, and the \citet{Tacchella2017} UV-$\beta$ measurement. $A_V$ profiles from this work are truncated at 0.2 $\mathit{SMA}_{\,\mathrm{med}}$, as indicated by the vertical coloured line segments, which is the typical limit of our resolution after PSF correction. Our profiles agree reasonably well with \citet{Tacchella2017}, who measure stellar continuum absorption, but disagree sharply with the H\,$\alpha$ result of \citet{Nelson2016}.}
		\label{fig:avcompare}
	\end{figure*}
	
	\label{sec:avcompare}
	As noted, relatively few previous studies have presented radial profile measurements of $A_V$ or sSFR  in the manner
	shown in this paper. However, two recent studies have appeared that permit a direct comparison.
	Fig.~\ref{fig:avcompare} compares our $A_V$ profiles to Balmer-based dust profiles using $\mathrm{H}\,\alpha/\mathrm{H}\,\beta$ from \citet{Nelson2016}, and to UV-based dust profiles using the slope $\beta$ of the UV continuum \citep{Meurer1999}, from \citet{Tacchella2017}. The Balmer decrement results have been interpolated to fit the same stellar mass bins as in our work; they measure dust extinction towards star-forming regions.  Our \textit{UVI} method, based on fits to optical/near-IR SEDs, measures continuum extinction to the average star.   The UV-$\beta$ method measures the extinction of the young-star stellar continuum and thus might be expected to be intermediate between the other two.
	
	In fact, Fig.~\ref{fig:avcompare} shows that the UV-$\beta$ method and the \textit{UVI} method yield similar radial trends in $A_V$, with highest extinction values at centres and mild decreases towards the galaxy outskirts by about a factor of 2 at 2 \textit{SMA}. A similar decline was found by \citet{Wuyts2012}, who also studied star-forming galaxies at $0.5<z<1.5$ and $\log M_\star/\mathrm{M}_{\sun}>10.0$, though no explicit radial profiles were shown. These mild trends also agree with local galaxies observed by \citet{Wild2011}, where continuum extinction was found to drop by only around 20--30 per cent from 0.35 to 1 Petrosian radius, more strongly in high-SFR galaxies.  
	
	In contrast, there is a strong discrepancy between both continuum methods in Fig.~\ref{fig:avcompare} and the $A_{\mathrm{H}{\,\alpha}}$ results from \citet{Nelson2016}, which exhibit sharp peaks in galaxy centres but nearly transparent profiles beyond 2 kpc.  This radial trend is very much steeper than the slow fall-offs in both continuum methods.  A popular two-component dust model (e.g., \citealt{Calzetti2000, Charlot2000, Wild2011}) permits $A_{\mathrm{H}\,\alpha}$ to be higher than the continuum attenuation due to extra attenuation toward star-forming regions.  However, measurements of both nearby galaxies and distant galaxies out to $z\sim1$ consistently find $A_{\mathrm{H}\,\alpha}/A_{\mathrm{cont}} \sim 2$ (e.g., \citealt{Calzetti2000, Wild2011, Wuyts2013}), at variance with the much larger ratio variations in Fig.~\ref{fig:avcompare}. Taking into account potential systematic uncertainties may mitigate this somewhat, such as [\ion{N}{ii}] contamination, inappropriate  SFH assumptions, and different stacking methods. However, it is highly unlikely that the full discrepancy can be explained by tweaks to the conventional picture.
	
	More likely, if low $A_{\mathrm{H}\,\alpha}$ in the outer parts of distant galaxies holds up with further data, this will signal a breakdown in our current geometrical model of dust, which is putting dust of all types into foreground screens.  More plausibly, dust and gas are mixed with stars, but in different ways for different stellar populations.  The foreground dust model for H$\,\alpha$ regions may also be wrong; the geometry of dust in \ion{H}{ii}  regions is complex and varies radially within the galaxy in at least one nearby object \citep{liu2013}.  Moreover, a major difference between local galaxies and distant galaxies is the much higher dust content in the latter, which averages 1.5--2.5 mag in the high-mass bins.  That much dust might possibly lead to weird  `skin effects' \citep{Wild2011} if certain populations lie farther in front of the dust than others.  Even so, constructing a model in which H$\,\alpha$ in dusty galaxies has essentially \emph{zero} attenuation, even in the outer regions, appears challenging.

	\subsection{sSFR profiles}  
	\label{sec:ssfrcompare}
\begin{figure*}
		\centering
		\includegraphics[width=3.7in]{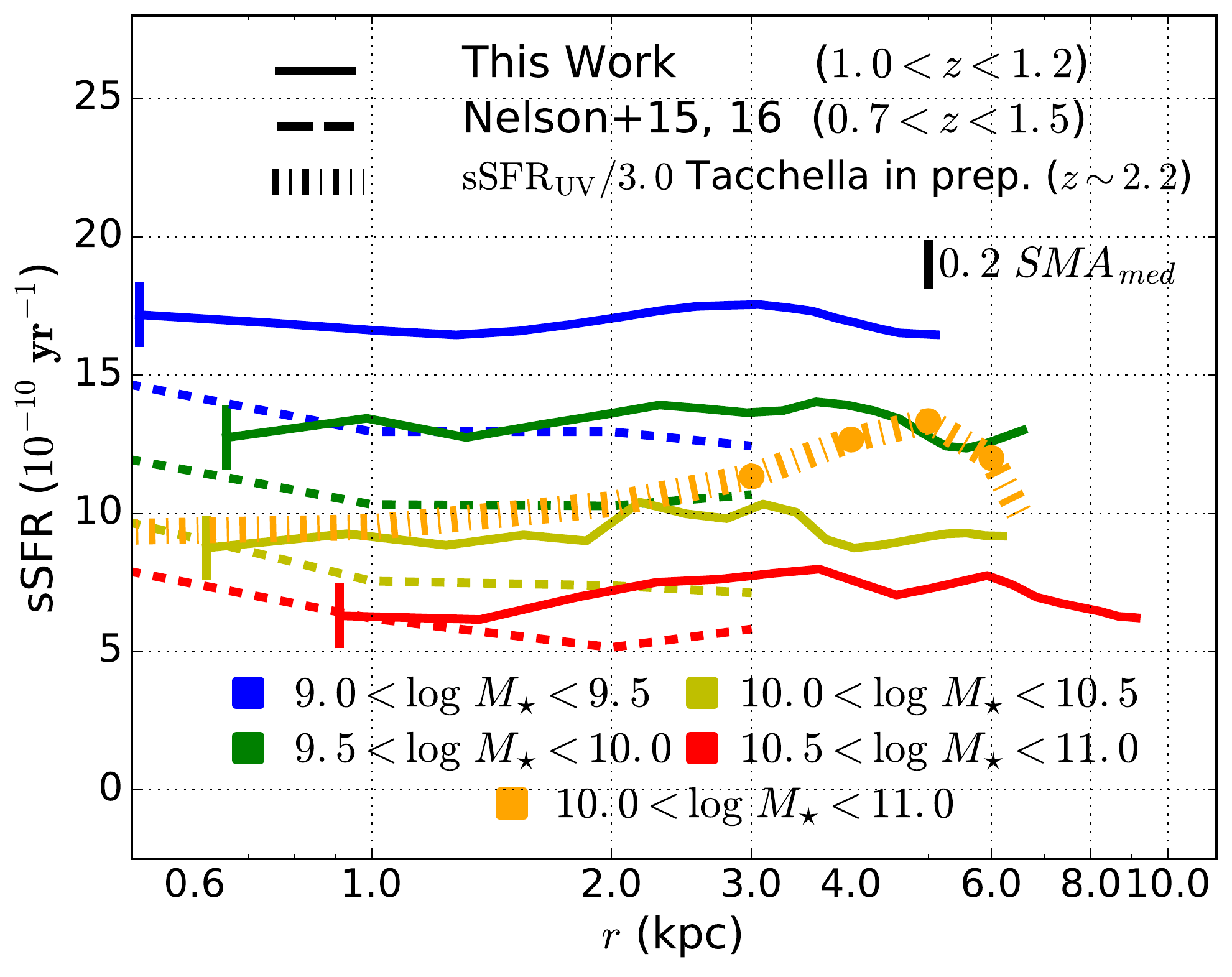}
		\caption{Dust-corrected sSFR profiles from the same three works as introduced in Fig.~\ref{fig:avcompare} (\citealt{Nelson2016}, \citet{Tacchella2017}, and this work), in combination with the 3D-HST H\,$\alpha$ map results by \citet{Nelson2016a}. As with Fig.~\ref{fig:avcompare}, sSFR profiles of this work are truncated at 0.2 $\mathit{SMA}_{\,\mathrm{med}}$.  The profiles shown in \citet{Nelson2016a} are corrected by the corresponding $A_V$ profiles from \citet{Nelson2016}. All profiles agree remarkably well, including the \citet{Nelson2016a, Nelson2016} result, whose corresponding $A_V$ profiles were deviant in Fig.~\ref{fig:avcompare}. See main text for details.}
		\label{fig:ssfrcompare}
	\end{figure*}

 Fig.~\ref{fig:ssfrcompare} compares sSFR profiles in different stellar mass ranges.
 In contrast to $A_V$, sSFR profiles are much more consistent. 
 
 The dashed lines again represent sSFR derived from H$\alpha$ emission in \citet{Nelson2016a}, corrected by the $A_{\mathrm{H}\,\alpha}$ profiles in Fig.~\ref{fig:avcompare} (which came from \citealt{Nelson2016}). In the correction, we assume the same differential attenuation law as  \citet{Wuyts2013}:
	\begin{equation}
		A_{\mathrm{extra}} (\lambda)=0.9 A_{\mathrm{cont}} (\lambda)-0.15 A_{\mathrm{cont}} (\lambda)^{2}
	\end{equation} \\ where $A_{\mathrm{extra}}=A_{\mathrm{\mathrm{H}\alpha}}-A_{\mathrm{cont}}$ is the additional attenuation of light at galaxy \ion{H}{II} regions, compared to the attenuation caused by ambient diffuse dust in other regions ($A_{\mathrm{cont}}$). We rewrite the equation above by replacing $A_{\mathrm{cont}}$ with $A_{\mathrm{\mathrm{H}\alpha}}$:
\begin{equation}
	A_{\mathrm{extra}} (\lambda)=A_{\mathrm{\mathrm{H}\alpha}}(\lambda)-6.33+\sqrt{40.11-6.67A_{\mathrm{\mathrm{H}\alpha}}(\lambda)}
\end{equation}
The sSFR profiles in \cite{Nelson2016a} are calculated from the quotient between H$\alpha$ and WFC3/F140W surface flux density (corresponding to rest-frame 7000 \AA\ at $z\sim1$, almost the same wavelength as  H$\alpha$). Therefore we include the corrections both from H$\alpha$ and from the broadband in the final sSFR calculation:
\begin{equation}
\begin{split} 
\mathrm{sSFR}_{\mathrm{corr}}&=\mathrm{sSFR}_{\mathrm{uncorr}}\cdot 10^{0.4A_{\mathrm{\mathrm{H}\alpha}}(6560\text{\AA})-0.4A_{\mathrm{cont}}(7000\text{\AA})}\\
&\simeq \mathrm{sSFR}_{\mathrm{uncorr}}\cdot 10^{0.4A_{\mathrm{\mathrm{H}\alpha}}(6560\text{\AA})-0.4A_{\mathrm{cont}}(6560\text{\AA})}
\\&=\mathrm{sSFR}_{\mathrm{uncorr}}\cdot 10^{0.4A_{\mathrm{extra}}(6560\text{\AA})}
\end{split}
\end{equation}    
 \\ where $\mathrm{sSFR}_{\mathrm{uncorr}}$ is the value presented by \citep{Nelson2016a}, and $\mathrm{sSFR}_{\mathrm{corr}}$ is the dust corrected sSFR.
 This correction more than removes the steep central drops in raw $A_{\mathrm{H}\,\alpha}$ seen in \citet{Nelson2016a}, showing that they are plausibly due to dust and not to drops in central sSFR. The orange dotted curve are the sSFR  values measured by \citet{Tacchella2017}  using the UV-$\beta$ method, which have been scaled down by a factor of 3 to correct for the general decline of sSFR with redshift (see \citealt{Rodriguez-Puebla2016}). Considering the vastly different radial dust corrections in \citet{Nelson2016} versus the two continuum methods, it is remarkable that all three methods wind up giving sSFR profiles that broadly agree in shape and overall magnitude. Generally flat sSFR profiles are seen, though a slight central fall of about (20 per cent) may be present in the most massive bin in our data (red curve). This is discussed below. 
	
	\section{Discussion}
	\label{sec:discussion}
	The previous sections have derived consistent trends in the radial
	behaviors of both sSFR and $A_V$ using the UVJ-based calibration,
	which in turn is founded on the conventional SED-fitting approach
	that assumes $\tau$-models, solar metallicity, the Calzetti attenuation 
	law, and dust in a foreground screen.
	Nevertheless, there are two puzzling aspects of the data that prompt further 
	discussion:
	
	1) A well established property of galaxies is that their stellar radii 
	grow with
	time \citep{VanDokkum2013, Patel2013a}.  
	In contrast, the present analysis finds that 
	specific star-formation rates are
	roughly the same at all radii.  If this were literally true at all 
	times and all
	radii, stellar radii would not grow, leading to a contradiction with
	the data.
	
	2) A second issue is the ratio of $A_V$ to stellar-mass surface
	density.  From Fig.~\ref{fig:avprofile}, we see that
	this ratio increases radially outward by a factor of  $\sim$10 over the 
	range of
	radii sampled.  As shown below, if dust scales roughly with gas, this 
	implies a large
	outward increase
	in gas-to-star density, which is surprising for galaxies that 
	supposedly have the same sSFR at
	all radii.  Alternatively, a large outward increase in the 
	dust-to-gas ratio could
	reconcile the data, but the magnitude of the needed effect is again large
	and is backwards from
	normal galaxy models, which tend to be metal-poor in their outer parts.
	
	The following sections discuss these two potential problems.  We 
	consider
	vulnerabilities in the conventional analysis that might help to explain away
	or mitigate the above results.
	We also compare to recent hydrodynamic models of galaxy formation, which, 
	as will be seen,
	are largely in accord with these observed properties.
	\begin{figure*}
		\includegraphics[width=5.9in]{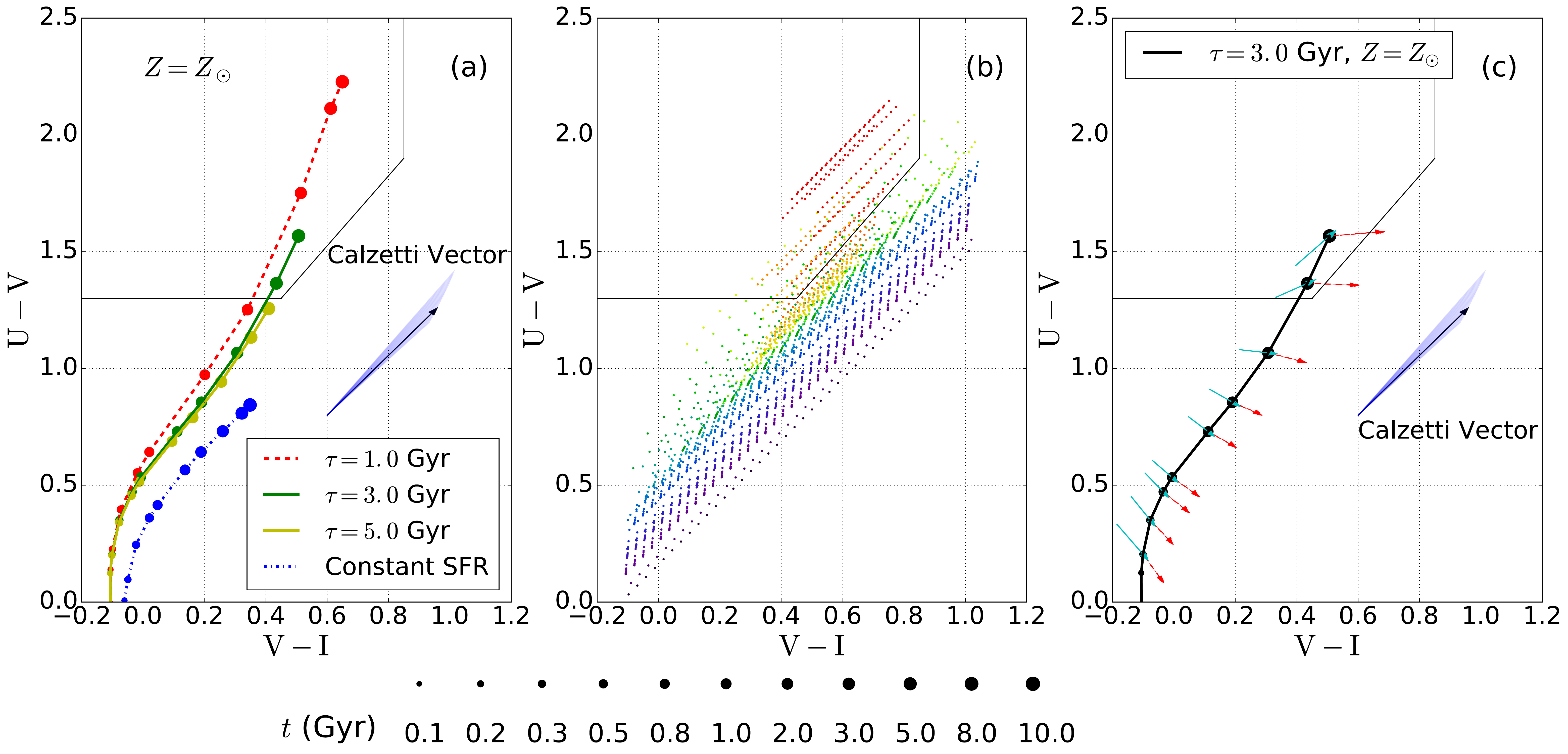}
		\caption{Panel a): Dust-free SPS tracks for three exponentially 
			declining SFHs [SFR$\propto \exp(-t/\tau)$] and a constant star-forming 
			SFH, all with solar metallicity.  Tracks are generated from the 
			\citet{Bruzual2003a} library with a \citet{Chabrier2003} IMF.  An offset 
			of 0.15 mag has been subtracted from model $V-I$ in order to locate the 
			tracks consistently relative to galaxies in \textit{UVJ} and \textit{UVI}; concerted 
			efforts to identify the cause of this offset failed, and the correction 
			has been applied empirically.   Black arrows show the dust reddening 
			vector and its variations determined by \citet{Salmon2016}, and points 
			with different sizes mark different stellar ages, as indicated at 
			bottom. The age effect has a steeper slope than the Calzetti vector on 
			the \textit{UVI} plane, except for the constant SFR case. Panel b): Model 
			distribution of points in the \textit{UVI} plane mixing galaxies with different 
			ages, exponential SFH time scales ($\tau$, 0.1--15.0 Gyr) and dust 
			extinction values on the \textit{UVI} diagram. Each point is generated by the SPS 
			models satisfying the conventional modeling assumptions and is 
			colour-coded by sSFR inferred from $\tau$ and age $t$.  Stripes of 
			constant sSFR along the Calzetti vector are clearly evident, and points 
			at a given location in \textit{UVI} all have similar sSFR despite their being a 
			mixture of ages and $\tau$'s. 
			Panel c): The effect of varying 
			metallicity on a $\tau=3.0$ Gyr, $Z=Z_{\sun}$ SPS track. Red dashed arrows 
			show the effect of increasing stellar metallicity from $Z_{\sun}$ to $2.5 Z_{\sun}$, and cyan solid lines show the effect of decreasing metallicity from $Z_{\sun}$ to $0.2 Z_{\sun}$. }
		\label{fig:spstrack}
	\end{figure*}
	
	\subsection{Critique of $A_V$ values}
	\label{sec:avcritique}
	We start the discussion of dust by reviewing how the conventional values 
	of $A_V$ from SED fitting are derived.
	This is explained by Fang et al.~(2017, submitted) and illustrated here in Fig.~\ref{fig:spstrack}.  Panel a) shows a variety of $\tau$-models [i.e., models with exponentially declining SFHs, $\mathrm{SFR(t)}\propto \exp(-t/\tau)$]
	without dust.  $\tau$-models generate tracks whose slopes 
	are steeper than the Calzetti vector, which permits separating their 
	stellar population ages from dust reddening. Adding dust to 
	$\tau$-models stretches them out along the Calzetti vector, producing 
	the familiar stripes of constant sSFR, with sSFR decreasing upwards.  
	This is illustrated in panel b), which mixes together solar-metallicity  
	$\tau$-models with different ages, $\tau$'s, and $A_V$ and colour-codes each 
	point by its sSFR.  It is seen that models near a given spot in \textit{UVI} have 
	remarkably similar sSFRs despite having a mixture of ages and 
	$\tau$'s. Note that this pattern is violated by the constant SFR model 
	(infinite $\tau$), which evolves \emph{parallel} to 
	the reddening vector at late times, despite having a falling sSFR.   
	sSFR  would be modestly overestimated for such populations using the 
	conventional SED method, while $A_V$ would be overestimated by up to 1 
	mag for periods of constant star formation lasting for several billion 
	years.  To derive $A_V$, a galaxy is slid backwards along the Calzetti 
	vector to intersect the $\tau$ locus, and the resulting $A_V$ varies 
	approximately linearly with $V-I$.

	We note that the
	conventionally derived $A_V$ values have strong support. For example, 
	UV-infrared excess (IRX) is an
	independent measure of dust besides the optical/near-IR 
	SED method used here.  IRX has been shown to correlate well with both 
	$r-K$ \citep{Arnouts2013} and with $V-J$ \citep{Forrest2016}, 
	supporting the notion that an optical/near-IR colour index can measure 
	dust.  The correlation between $V-I$ and $V-J$ established in this paper 
	then validates $V-I$. Also \citet{Tacchella2017} use the relatively independent UV-$\beta$ method, 
	and our agreement with them in Fig.~\ref{fig:avcompare} is further 
	support.
	
\begin{figure*}
		\centering
		\includegraphics[width=4in]{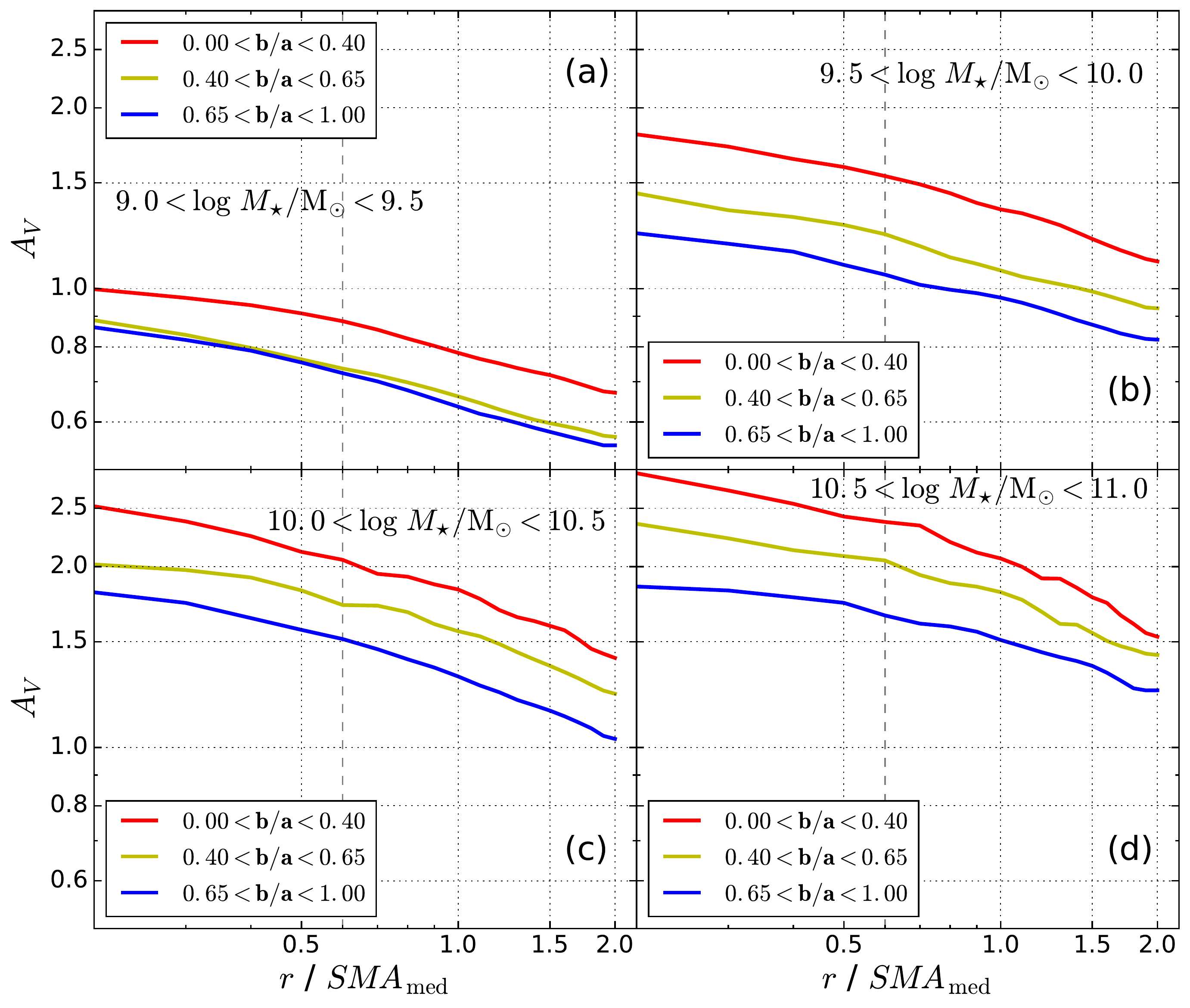}
		\caption{ $A_V$ profiles as a function of axis ratio $b/a$ and 
			galaxy stellar mass. The data are stacked as in Fig.~\ref{fig:UVIgradient}, but edge-on galaxies are now included.  For all 
			stellar-mass bins, objects with smaller axis ratio have higher dust 
			extinction, as expected for edge-on disks. Vertical axes are plotted on 
			a logarithmic scale to illustrate relative radial changes.  Nearly 
			constant vertical shifts indicate that $A_V$ profiles maintain shape and 
			shift by a constant factor at all radii as inclination varies.}
		\label{fig:av_ba}
	\end{figure*}

	As a further check, we have investigated the dependence of $A_V$ on 
	galaxy inclination.  Star-forming galaxies are beginning to settle to 
	rotating disks by $z\sim1$ \citep{Kassin2012}, and, if $A_V$ measures 
	dust, it should be higher in edge-on galaxies (see study of local disk galaxies by \citealt{Devour2016}).  A test of this is shown 
	in Fig.~\ref{fig:av_ba}, which shows $A_V$ profiles of galaxies grouped by inclination. Flattened 
	galaxies are now included, and the sample is divided into bins of mass 
	and axial ratio.  It is seen that $A_V$ rises with inclination, as 
	predicted. The \emph{relative} change in $A_V$ with radius is 
	similar for all inclinations, as shown by the constant visual offset 
	between the (however, logarithmically plotted) curves for different axis-ratio 
	samples. This indicates a similar radial \emph{pattern} of dust at different 
	inclinations. Axis ratio data thus further support the scenario that $V-I$ 
	measures dust and that $V-I$ colour gradients mainly reflect dust gradients.
	
	A final comparison is with the recent hydrodynamic models of galaxy 
	formation by \citet{Ceverino2014} (also see \citealt{Zolotov2015}), as analyzed by 
	\cite{Tacchella2016a}. For low-mass galaxies with $\log M_\star/\mathrm{M}_{\sun} < 10.2$ near 
	$z = 1$, their fig.~7 shows a radial decline in stellar mass surface 
	density, $\Sigma_\star$, of $\sim 2$ dex/dex over our range of measured radii, 
	compared to a fall in gas surface density, $\Sigma_{g}$, of only $\sim0.3$ dex/dex.  This gradient ratio 
	of $\sim$0.15 in gas to stars agrees well with the ratio $\mathrm{d} \log A_V/\mathrm{d} 
	\log \Sigma_\star$ = 0.25 seen in our data in Fig.~\ref{fig:avprofile}, 
	assuming that $A_V$ scales in proportion to $\Sigma_{g}$.  A similar 
	ratio is seen for their high-mass galaxies though both declines are 
	individually larger. Thus our data look like models.
	
	In short, continuum data from several
	studies (\citealt{Wuyts2012}, \citealt{Tacchella2017}, and this
	work) indicate slowly falling $A_V$ profiles in the outer parts of distant galaxies, and thus the
	presence of considerable gas there if the dust-gas ratio is approximately constant \citep{Rachford2009, Bolatto2013, Sandstrom2013}. 
		
	We note in passing that, although both data and theory support 
	the conventional SED-fitting method, which employs $\tau$-models, 
	$A_V$ values could be lower using different 
	stellar population models.  An experiment with a toy composite stellar 
	population is shown in Appendix \ref{sec:uvi_composite}, formed by combining different 
	amounts of very old and very young stars.  Such a mixture 
	might be appropriate to a fully quenched bulge surrounded by a strongly 
	star-forming outer disk. The total central quenching in this model is 
	thus too extreme for our galaxies, but they may be evolving in this 
	direction. It is seen that the effect of compositeness is to 
	create models on curved loci that lie to the right of the $\tau$-model 
	tracks, and thus yield lower values of $A_V$.  A second example is the 
	constant star-formation model (infinite $\tau$) in Fig.~\ref{fig:spstrack}, which also evolves off to the right.  Our point is 
	that using $\tau$-models is not a neutral assumption about $A_V$ but rather tends to 
	maximize it.  This should be kept in mind, especially at late times, 
	when compositeness tends to increase.
	
	Furthermore, alternative dust geometry models may also change our understanding of the dust surface density distribution. Different from the dust screen model assumed in this work, which predicts a proportional relation between SED-fitted $A_V$ and dust surface density, $A_V$ can be a highly non-linear function of dust surface density if in reality dust is homogeneously mixed with stars (``dust slab" model, see eq.~3 of \citealt{Genzel2013}). Similar models have been considered in several local studies (e.g.~\citealt{Tuffs2004, Wild2011, Genzel2013, Devour2016}), however still in question for high redshift galaxies \citep{Wuyts2011, Nordon2013, Salmon2016}. As a verification of the non-linear effect, we have used \textsc{fast} to fit model SEDs of SFMS galaxies, created assuming exponential declining SFHs, solar metallicity and different amount of attenuation following the slab dust model. To best resemble the real data and SED fitting process by \citet{Santini2015}, these model galaxies are in the same mass and redshift range as the observations, and their SEDs are sampled in the same observed bands as the GOODS-South photometry catalog \citep{Guo2013} to serve as the SED-fitting input. We find that when the output $A_V$ declines from $\sim$2.0 mag to $\sim$1.0 mag, which is the typical case in Fig.~\ref{fig:avprofile}, the dust surface density (proportional to the total optical depth) drops by a factor of 6. This value is still smaller than the observed stellar mass density change (a factor of 10-20). Moreover, \citet{Nordon2013} suggested that the realistic dust-star geometry of high redshift galaxies should be a hybrid of the dust screen model and the slab model, which indicates that the factor derived here with the slab model is an upper limit.

	In summary, we find good agreement of $A_V$ profile measurements with other observation results and hydrodynamic simulations, assuming that the dust-gas ratio is approximately constant. When trying to infer dust (gas) surface density from $A_V$ profiles, the slab dust model predicts a 3 times steeper drop than the screen model, which indicates that dust geometry can significantly influence our understanding of the dust surface density distribution. However, this only mitigates but does not remove the even larger difference between the inferred radial dust (gas) density slope and the stellar density slope. Such persisting difference is also found to exist in hydrodynamic simulations \citep{Tacchella2016a}. The deviation from $\tau$ SFH could bring bias to our measured $A_V$ value, but will also break our established consistence with other independent works.
	
	The next section discusses this result in comparison to more steeply falling
	SFR profiles.
	
	\subsection{Critique of sSFR values}
	\label{sec:ssfrcritique}
	The second problem with the conventional SED-fitting approach is that 
	the resulting sSFR profiles appear to be constant with radius.  If 
	true at all times and all radii, this would yield star-forming galaxies with 
	constant stellar-mass effective radii.  In contrast, light-weighted 
	effective radii are seen to increase roughly as $M_\star^{0.3}$ (\citealt{Patel2013a, VanDokkum2013}, 
	see also \citealt{Papovich2015}). It is true that 
	mass-weighted and light-weighted radii can differ in galaxies with 
	stellar-population gradients \citep{Fang2013a}, but galaxies with flat 
	sSFR profiles should not have such gradients.  A 
	related datum is the evolution of galaxies in ($\Sigma_1$, $M_\star$) space, 
	which plots stellar mass surface density within 1 kpc versus total 
	stellar mass.  $\Sigma_1$ appears to grow more slowly than $M_\star$ 
	\citep{Barro2015}, again indicating lower central stellar-mass growth.
	
	We turn again to hydrodynamic models for help in reconciling these 
	discrepancies.  We find that plots of model sSFR in fig.~7 of \citet{Tacchella2016a} 
	are remarkably flat over our measured range of radii, especially in 
	smaller galaxies at early times.  The resemblance to our data is 
	striking.  Nevertheless, the stellar-mass radii of these same galaxies 
	are growing.  This is because gas is being added \emph{beyond} our measured 
	radii, where considerable star formation is taking place.  Hence, our 
	decision to limit the data to 2 $R_e$ may have inadvertently eliminated 
	regions that are contributing most to radius and stellar-mass growth.  
	Taking this limitation into account would reconcile our flat sSFR profiles 
	with the observed radii growth of star-forming galaxies and also with 
	the slower growth of $\Sigma_1$ relative  to $M_\star$ in 
	\citet{Barro2015}. In essence, galaxies grow inside-out in an overall sense while star-forming, but our data 
	only sample the inner regions. 
	
	Beyond this, the streaming of stellar mass on to galaxy outskirts via minor mergers and accretion
	may also contribute to the galaxy size growth, which has been discussed in recent works \citep{Welker2017}. 
	We leave this possibility open considering that the roles of mergers in galaxy formation 
	is still in debate (e.g., \citealt{Lotz2010,Puech2014} and references therein). 
	
	We further note that, although sSFR is flat, SFR falls steeply (see 
	Fig.~\ref{fig:avprofile}) in contrast to gas and/or $A_V$, which
	both fall slowly.  This behavior is a feature not only of
	the data but also of the hydrodynamic models.  A steep fall in SFR
	plus a slow fall in gas  is not consistent with the Kennicutt law 
	\citep{Kennicutt1992}, which says that $SFR\propto \Sigma_g^{n}$, 
	where $n \sim 1.4$.  Obtaining agreement with the models and our data would 
	require increasing $n$ to $\sim 3$--4 over our measured range of radii. Local spatially resolved galaxies exhibit the same trend, and suggest that
	the high value of $n$ at galaxy outskirts is caused by \ion{H}{i} dominating over $\mathrm{H}_2$ \citep{Bigiel2008}.
	This scenario should be pursued at high redshift in future measurements.
	
	Even though our data do not probe large radii,
	they \emph{are} a good probe of inner quenching.  A major 
	finding of this paper is that, as an average trend bulges are not strongly quenched over our measured range of masses and 
	redshifts (cf. the empty quenched 
	regions in Fig.~\ref{fig:UVIgradient_re} and \ref{fig:UVIgradient}).  
	We now attempt to assess the levels of central star formation more 
	quantitatively.  The sSFR profiles in Fig.~\ref{fig:ssfrprofile} are 
	uniformly flat for all lower masses but may show central dips in the 
	highest mass bin.  Fitting sSFR as a power of $r$ to these 
	profiles yields an average deficit of sSFR at 0.2 \textit{SMA} relative to 
	1.0 \textit{SMA} of 20 $\pm$5 per cent, increasing to 24$\pm$5 per cent in the lowest redshift 
	bin ($z = 0.4$--0.8).  In other words, sSFR profiles are flat at low mass 
	but show small central depressions at high mass, possibly growing 
	modestly with time. Bulge quenching thus appears to set in near $z \sim 
	1$ at high masses, evolving ultimately to the sequence of strongly 
	quenched bulges in high stellar-mass galaxies that is seen today (e.g., 
	\citealt{Delgado2015,Delgado2016}).
	
	This picture broadly agrees with previous work on $z\sim1$ galaxies, 
	with some differences.  \citet{Wuyts2013}
	used dust-corrected H\,$\alpha$ to measure radial star-formation patterns 
	in massive galaxies above $\log M_\star/\mathrm{M}_{\sun} = 10.0$ at $z = 
	0.7$--1.5.  The main result was constant sSFR throughout most of the 
	galaxies, in agreement with our flat sSFR profiles.  Some reduction in 
	central sSFR was noted, but no numbers were given. \cite{Patel2013a} 
	measured the time evolution of stellar mass profiles in Milky Way 
	progenitors and found that growth rates inside and outside 2 kpc were 
	the same down to $z = 0.9$ (when corrected for $M/L$ differences) and 
	diverged slowly after that.  In a similar study, \cite{VanDokkum2013} 
	found identical inner and outer growth rates down to $z = 0.6$.
	\citet{Shibuya2015} found a steep increase of S\'{e}sic index for star forming galaxies at $z\sim 1$ and $10.5<\log\ {M}_{\star}/\mathrm{M_{\sun}}<11.0$, compared to the nearly invariant trends in other mass-redshift ranges. The onset of bulge quenching is detectable in our most massive bin, which is closest to the Milky Way progenitor mass at $z\sim 1$ ($\sim 10.5\ \mathrm{M}_{\sun}$), in a rough agreement with the two Milky Way progenitor studies, and lies in exactly the same range where S\'{e}sic index is seen to increase.
	
	In a study using CANDELS images, \citet{Liu2016} measured colour 
	gradients in star-forming galaxies at $z\sim 1$. This study used the same radial photometry catalog used in the present study. Radial rest-frame $NUV-B$ 
	colour gradients were nearly zero for $\log M_\star/\mathrm{M}_{\sun} <10.0$ after 
	dust correction, but significant residual colour gradients were found 
	above this, corresponding to $\mathrm{d} \log \mathrm{sSFR}/\mathrm{d} \log r\simeq 0.4$ at $10.5<\log M_\star/ \mathrm{M}_{\sun}<11.0$.  
	The mass trend agrees with our results, but the central fall of 
	$\sim$60 per cent in high-mass galaxies is three times larger than what we find. 
	The \citet{Liu2016} study determined dust and SFR by running \textsc{fast} directly on B 
	through H band \emph{HST} images, whereas our study attempts to 
	match a calibration based on SEDs that originally includes IRAC bands.  Preliminary tests 
	fitting \textsc{fast} with and without IRAC show significant scatter and biases, 
	which may contribute to the difference.
	
	The only previous work in potentially serious disagreement with the present results 
	is \citet{Tacchella2015}, which used H$\alpha$ fluxes to measure sSFR 
	gradients in star-forming galaxies at  $z\sim 2.2$.  In massive galaxies 
	with $\log M_\star/\mathrm{M}_{\sun}=10.8$--11.7, central sSFR was found to be only 
	$\sim$1/10 of outer sSFR in the same objects (cf. their fig.~1B(c)).  However, 
	no information on the H$\alpha$/H$\beta$ ratio as a function of position 
	was available.  The subsequent data of \citet{Nelson2016} in Fig.~\ref{fig:avcompare} show that central H\,$\alpha$ can be strongly 
	absorbed, and correcting that yields flat sSFR profiles 
	(cf. Fig.~\ref{fig:ssfrcompare}). The integrated dust attenuation value $A_V$ of the massive galaxies studied by \citet{Tacchella2015} is around 1.0 mag, which indicates that $A_V$ in the galaxy centers can be even larger than that value. So radial dust corrections are needed to precisely constrain sSFR profiles. Subsequent work by \citet{Tacchella2017} on $z \sim 2.2$ galaxies used the UV-$\beta$ technique and found 
	flat sSFR profiles for galaxies in the range $\log M_\star/\mathrm{M}_{\sun} =10.0$--11.0, 
	as shown in Fig.~\ref{fig:ssfrcompare}. Similar measurements at $z\simeq 4$ found centrally depressed sSFR for galaxies with high central mass density, but no confirmed central sSFR drop is seen when their sample is binned into stellar mass bins \citep{Jung2017}.  
	
	We conclude this section with a discussion of other uncertainties that 
	may affect our sSFR profiles. \citet{Salmon2016} conducted an 
	independent study of the attenuation law and found good overall agreement 
	with Calzetti but a small change in slope that varies with the amount of 
	attenuation, producing a curved reddening line in \textit{UVJ} (and \textit{UVI}) above the 
	Calzetti vector.  Using the Salmon law would alter sSFR gradients by 
	typically 0.1 dex over our measured radii, by amounts that depend on 
	dust content.  Low-mass, low-dust galaxies would have central sSFR 
	\emph{increased} by 0.1 dex, while high-mass high-dust galaxies would 
	have  central sSFR \emph{decreased} by 0.1 dex, increasing central quenching.  
	
	Metallicity gradients, if present, would further impact sSFR profiles.
	Panel c) of Fig.~\ref{fig:spstrack} shows the effect of  varying $Z$ on a 
	galaxy's location in the \textit{UVI} diagram.  Broadly, increasing $Z$ shifts 
	points to the right and (possibly) down.  If stars are more metal-rich 
	at the centres of galaxies, we would first move that point left (and 
	upwards) to correct it to solar metallicity and then apply the present 
	\textit{UVI} calibration.  In the presence of metallicity gradients, this 
	correction would move central points more than outer points, creating a 
	tilt with respect to the Calzetti vector, and thus an sSFR gradient.  
	Higher central metallicities would \emph{reduce} central sSFR's, and thus 
	produce more bulge quenching than the conventional method, which  uses uniform 
	solar metallicities.
	
	The size of the resulting gradients depends on metallicity, since stellar 
	colours are less sensitive at low $Z$ (see Fig.~\ref{fig:spstrack}c).  A 
	gradient of $\mathrm{d} \log Z/\mathrm{d} \log r = -0.6$ dex/dex produces $\mathrm{d} \log \mathrm{sSFR}/\mathrm{d} 
	\log r = +0.2$ dex/dex at super-solar metallicities, but the effect is 
	three times smaller at sub-solar metallicities.  IFU surveys of distant galaxies
	\citep{Stott2014, Wuyts2016, Leethochawalit2016} 
	indicate that gas-phase metallicity gradients mainly cluster in the 
	range -0.05 to +0.05 dex/kpc  at $z\approx 0.5$--2.5, which converts to 
	approximately -0.5 to +0.5 dex/dex for our galaxies. However, galaxies that are
	initiating bulge quenching would tend to be more massive and more highly differentiated 
	radially, and thus might be expected to have metallicity gradients near 
	the upper end of this range.  
	
	We have thus identified two corrections -- metallicity gradients
	and corrections to the attenuation law -- that both tend to increase 
	the modest central sSFR depressions that we detect.  As metallicity cannot be 
	measured from \textit{UVI} alone, correcting for it will require detailed 
	absorption-line spectra as a function of radius. Putting
	all numbers together, we can imagine that the raw central drops 
	of 20--25 per cent measured 
	in the massive bin  could be perhaps twice as  large.  However, none
	of these factors can actually move raw points into the quenched region in \textit{UVI}, 
	and hence our averaged galaxy centres are not close to being fully quenched. 
	
	The final factor to be considered is the accuracy of stellar 
	population models.  As long as stellar populations are $\tau$-models or 
	mixtures of $\tau$-models, our sSFR values are accurate, as shown in 
	panel b) of Fig.~\ref{fig:spstrack}.  Section \ref{sec:avcritique} mentioned
	two types of models that evolve off the $\tau$ locus to the right: 
	composite models, as shown in Fig.~\ref{fig:composite} and discussed in 
	the Appendix \ref{sec:uvi_composite}, and the model with constant SFR (infinite $\tau$) in  
	Fig.~\ref{fig:spstrack}.  sSFRs for both of these models using the 
	present \textit{UVI} calibration would be overestimated.  However, as long as 
	stellar populations are reasonably uniform \emph{throughout} a galaxy, this 
	error does not affect gradients. A non-uniform attenuation model is also considered
	 in the Appendix \ref{sec:uvi_composite}, i.e., mixing dusty young stars
	with dust-free old stars, which could be closer to the realistic case of composite stellar population.
	But still, quenched centres cannot be recovered under such model tests.  This overall means our findings of flat radial sSFR trends
	are robust against straightforward modeling of composite stellar population.  
	But we acknowledge the subtle possibility that more sophisticated 
	stellar population models in the future, including varying stellar metallicity and more complex compositeness,
	can potentially alter the conclusion.

	To summarize, our measurements suggest that flat sSFR profiles are the norm 
	for low-mass galaxies but that bulge quenching sets in around $z \sim 1$ 
	at high masses.  Both findings are in general accord with previous works. 
	The apparent contradiction between flat sSFR profiles and radius growth 
	can be reconciled by noting that our data are truncated at 2 $R_e$ and 
	stars may be forming outside this radius, as seen in recent hydrodynamic 
	models.
	
	\section{Conclusions}
	\label{sec:conclusion}
	
	We have shown that measuring $A_V$ profiles is a promising way to study the
	interstellar medium and its distribution in distant galaxies.  The major conclusions
	from this paper are as follows:
	
	1) We analyze the radial colour gradients of star-forming galaxies from
	$z = 0.4$ to 1.4 and $\log M_\star/\mathrm{M}_{\sun} = 9.0$--11.0. \emph{HST}  images of CANDELS galaxies
	from $B$ through $H$ are used to derive rest-frame values of $U-V$ and $V-I$
	in annular bins. Galaxies are stacked in bins of mass and redshift, and radial colour 
	gradients for each bin are plotted in \textit{UVI}. The data cover $\sim$1 dex in radius 
	starting near 1 kpc  and extend to 2 $R_e$, where they are truncated.  
	
	2) The sizes of colour gradients increase with global $A_V$ and 
	with galaxy stellar mass.  Gradients
	parallel the Calzetti reddening vector in the \textit{UVI} plane, indicating that most of the gradient
	is due to dust, not sSFR.  Corrections
	are derived for PSF smearing, and the data are stacked in both arcsec and
	fractional effective radius, but the basic conclusions remain 
	the same.  In
	particular, galaxy centres averaged in redshift and stellar mass bins lie within the star-forming region of \textit{UVI} and are not quenched.
	
	3) The integrated colours of galaxies in the \textit{UVI} and \textit{UVJ} diagrams are 
	compared,
	and it is shown that \textit{UVI} can also be used to derive sSFR and $A_V$ 
	analogous to \textit{UVJ}.
	The calibrations for sSFR and $A_V$ derived by Fang et al.~(2017, submitted) for \textit{UVJ} 
	are transferred to \textit{UVI} using the integrated colours of galaxies in 
	common.  The 
	calibrations are based on fits to CANDELS catalogue values of sSFR and $A_V$ 
	that are derived using SED fits to
	integrated optical/near-IR photometry, assuming declining-$\tau$ stellar-populations, stellar metallicity, 
	a Calzetti attenuation law, and dust in a foreground screen.  The resulting values are similar 
	or identical
	to conventional values of sSFR and $A_V$ derived by using the \textsc{fast} SED 
	fitting code.  After mapping on to the \textit{UVI} diagram, 
	the values of sSFR and $A_V$
	are recovered with an rms error of 0.15 dex and 0.18 mag respectively.
	
	4) Integrated sSFR values decline weakly with both mass and time, while
	integrated $A_V$  increases
	strongly with mass. Integrated $A_V$ increases weakly with time at fixed 
	mass, but this
	trend reverses at the highest masses, where $A_V$ decreases with time, possibly due
	to a loss of ISM at late times.  $A_V$ is higher in edge-on galaxies, and 
	$A_V$ gradients
	are steeper in dustier galaxies.
	
	5) Radial sSFR and $A_V$ profiles based on our conventional
	SED fitting calibration are presented versus radius.  The sSFR profiles are
	generally flat except for the $\log M_\star/\mathrm{M}_{\sun} = 10.5$--11.0 bin, where values at 0.2 \textit{SMA}
	average 20 per cent smaller than at 1.0 \textit{SMA}, increasing to 25 per cent in
	the lowest redshift bin.  These trends are consistent with bulge quenching
	setting in near $z\sim1$ at high mass.  $A_V$ profiles fall smoothly by about a
	factor of two from the centre to the edge of the data at all masses
	and redshifts.
	
	6) Two puzzling aspects of the data are considered.  The first is that
	sSFR profiles are flat in most mass bins yet the radii of star-forming
	galaxies are observed to increase as $R_e\sim M_\star^{0.3}$.  This discrepancy
	can be explained if new gas and stars are being added outside
	our range of measured radii that we are not sensitive to.  A second point
	is that dust (gas) surface density declines much more slowly than
	stellar-mass surface density (assuming that $A_V$ is
	proportional to gas surface density and non-varying dust-to-gas ratio). Switching from the dust screen assumption to the slab model only mitigates but does not remove this difference. Comparing this to SFR
	profiles then implies a Kennicutt law with exponent $\sim$4, much steeper
	than the conventional value of 1.4.  Interestingly, both of these 
	features -- flat inner sSFR profiles and shallow radial declines in gas surface density --
	are also seen in recent hydrodynamic models, which also have growing radii, 
	flat inner sSFR profiles, and shallower
	fall-offs in gas than in stars \citep{Tacchella2016a}.

\section*{Acknowledgements}
We thank the anonymous referee for constructive comments. We acknowledge the CANDELS team members for the substantial effort in science operation and catalogue construction, including photometry, morphology, redshift, rest-frame photometry, and physical parameter works, which this work is heavily based on. WW thanks Shude Mao for undergraduate advising, Fengshan Liu and Sandro Tacchella for inspiring discussions and sharing pioneering results, Eric Bell for useful comments, and Tsinghua Center for Astrophysics, and University of California, Santa Cruz for generous supports. WW and SAK acknowledge support from NASA grant 15-ADAP15-0186, `What is the Dominant Mode of Star-Formation as a Function of Galaxy Mass and Redshift?'.  SMF, YG, DCK, and JJF acknowledge the NASA HST grant GO-12060.10-A and NSF grant AST-0808133. The aperture photometry project was supported by the NSFC (11103013 and 11573017). ZC acknowledges support from NFSC funding 11403016.

This work is based on observations taken by the CANDELS Multi-Cycle Treasury Program with the NASA/ESA HST, which is operated by the Association of Universities for Research in Astronomy, Inc., under NASA contract NAS5-26555. Also this work is based in part on observations made with the Spitzer Space Telescope, which is operated by the Jet Propulsion Laboratory, California Institute of Technology under NASA contract 1407. We also thank various important ground-based observation projects, which greatly empower the CANDELS catalogues by expanding observation data to ranges from UV to far IR. This research made use of {\sc Astropy}, a community-developed core Python package for Astronomy (Astropy Collaboration, 2013).
	
	\bibliographystyle{mnras}
	\bibliography{wang2017} 

\begin{thebibliography}{}
\makeatletter
\relax
\def\mn@urlcharsother{\let\do\@makeother \do\$\do\&\do\#\do\^\do\_\do\%\do\~}
\def\mn@doi{\begingroup\mn@urlcharsother \@ifnextchar [ {\mn@doi@}
  {\mn@doi@[]}}
\def\mn@doi@[#1]#2{\def\@tempa{#1}\ifx\@tempa\@empty \href
  {http://dx.doi.org/#2} {doi:#2}\else \href {http://dx.doi.org/#2} {#1}\fi
  \endgroup}
\def\mn@eprint#1#2{\mn@eprint@#1:#2::\@nil}
\def\mn@eprint@arXiv#1{\href {http://arxiv.org/abs/#1} {{\tt arXiv:#1}}}
\def\mn@eprint@dblp#1{\href {http://dblp.uni-trier.de/rec/bibtex/#1.xml}
  {dblp:#1}}
\def\mn@eprint@#1:#2:#3:#4\@nil{\def\@tempa {#1}\def\@tempb {#2}\def\@tempc
  {#3}\ifx \@tempc \@empty \let \@tempc \@tempb \let \@tempb \@tempa \fi \ifx
  \@tempb \@empty \def\@tempb {arXiv}\fi \@ifundefined
  {mn@eprint@\@tempb}{\@tempb:\@tempc}{\expandafter \expandafter \csname
  mn@eprint@\@tempb\endcsname \expandafter{\@tempc}}}

\bibitem[\protect\citeauthoryear{Arnouts et~al.,}{Arnouts
  et~al.}{2013}]{Arnouts2013}
Arnouts S.,  et~al., 2013, A{\&}A, 558, A67

\bibitem[\protect\citeauthoryear{Bakos, Trujillo  \& Pohlen}{Bakos
  et~al.}{2008}]{Bakos2008}
Bakos J.,  Trujillo I.,   Pohlen M.,  2008, ApJ, 683, L103

\bibitem[\protect\citeauthoryear{Barro et~al.,}{Barro et~al.}{2015}]{Barro2015}
Barro G.,  et~al., 2015, preprint (arXiv:1509.00469)

\bibitem[\protect\citeauthoryear{Barro et~al.,}{Barro
  et~al.}{2016}]{Barro2016a}
Barro G.,  et~al., 2016, ApJ, 827, L32

\bibitem[\protect\citeauthoryear{Bertin \& Arnouts}{Bertin \&
  Arnouts}{1996}]{Bertin1996}
Bertin E.,  Arnouts S.,  1996, A{\&}AS, 117, 393

\bibitem[\protect\citeauthoryear{Bigiel, Leroy, Walter, Brinks, de Blok, Madore
   \& Thornley}{Bigiel et~al.}{2008}]{Bigiel2008}
Bigiel F.,  Leroy A.,  Walter F.,  Brinks E.,  de Blok W. J.~G.,  Madore B.,
  Thornley M.~D.,  2008, ApJ, 136, 2846

\bibitem[\protect\citeauthoryear{Bolatto, Wolfire  \& Leroy}{Bolatto
  et~al.}{2013}]{Bolatto2013}
Bolatto A.~D.,  Wolfire M.,   Leroy A.~K.,  2013, ARA\&A, 51, 207

\bibitem[\protect\citeauthoryear{Brammer, van Dokkum  \& Coppi}{Brammer
  et~al.}{2008}]{Brammer2008}
Brammer G.~B.,  van Dokkum P.~G.,   Coppi P.,  2008, ApJ, 686, 1503

\bibitem[\protect\citeauthoryear{Brammer et~al.,}{Brammer
  et~al.}{2009}]{Brammer2009}
Brammer G.~B.,  et~al., 2009, ApJ, 706, L173

\bibitem[\protect\citeauthoryear{Brammer et~al.,}{Brammer
  et~al.}{2012}]{Brammer2012}
Brammer G.~B.,  et~al., 2012, ApJS, 200, 13

\bibitem[\protect\citeauthoryear{Brinchmann, Charlot, White, Tremonti,
  Kauffmann, Heckman  \& Brinkmann}{Brinchmann et~al.}{2004}]{Brinchmann2004}
Brinchmann J.,  Charlot S.,  White S. D.~M.,  Tremonti C.,  Kauffmann G.,
  Heckman T.,   Brinkmann J.,  2004, MNRAS, 351, 1151

\bibitem[\protect\citeauthoryear{Bruzual \& Charlot}{Bruzual \&
  Charlot}{2003}]{Bruzual2003a}
Bruzual G.,  Charlot S.,  2003, MNRAS, 344, 1000

\bibitem[\protect\citeauthoryear{Calzetti, Kinney  \&
  Storchi-Bergmann}{Calzetti et~al.}{1994}]{Calzetti1994}
Calzetti D.,  Kinney A.~L.,   Storchi-Bergmann T.,  1994, ApJ, 429, 582

\bibitem[\protect\citeauthoryear{Calzetti, Armus, Bohlin, Kinney, Koornneef  \&
  Storchi-Bergmann}{Calzetti et~al.}{2000}]{Calzetti2000}
Calzetti D.,  Armus L.,  Bohlin R.~C.,  Kinney A.~L.,  Koornneef J.,
  Storchi-Bergmann T.,  2000, ApJ, 533, 682

\bibitem[\protect\citeauthoryear{Ceverino, Klypin, Klimek, Trujillo-Gomez,
  Churchill, Primack  \& Dekel}{Ceverino et~al.}{2014}]{Ceverino2014}
Ceverino D.,  Klypin A.,  Klimek E.~S.,  Trujillo-Gomez S.,  Churchill C.~W.,
  Primack J.,   Dekel A.,  2014, MNRAS, 442, 1545

\bibitem[\protect\citeauthoryear{Chabrier}{Chabrier}{2003}]{Chabrier2003}
Chabrier G.,  2003, PASP, 115, 763

\bibitem[\protect\citeauthoryear{Charlot \& Fall}{Charlot \&
  Fall}{2000}]{Charlot2000}
Charlot S.,  Fall S.~M.,  2000, ApJ, 539, 718

\bibitem[\protect\citeauthoryear{D'Souza, Kauffman, Wang  \& Vegetti}{D'Souza
  et~al.}{2014}]{D'Souza2014}
D'Souza R.,  Kauffman G.,  Wang J.,   Vegetti S.,  2014, MNRAS, 443, 1433

\bibitem[\protect\citeauthoryear{Dahlen et~al.,}{Dahlen
  et~al.}{2013}]{Dahlen2013}
Dahlen T.,  et~al., 2013, ApJ, 775, 93

\bibitem[\protect\citeauthoryear{Devour \& Bell}{Devour \&
  Bell}{2016}]{Devour2016}
Devour B.~M.,  Bell E.~F.,  2016, MNRAS, 459, 2054

\bibitem[\protect\citeauthoryear{Dunlop et~al.,}{Dunlop
  et~al.}{2017}]{Dunlop2016}
Dunlop J.~S.,  et~al., 2017, MNRAS, 465, in press

\bibitem[\protect\citeauthoryear{Elbaz et~al.,}{Elbaz et~al.}{2007}]{Elbaz2007}
Elbaz D.,  et~al., 2007, A{\&}A, 468, 33

\bibitem[\protect\citeauthoryear{Elmegreen \& Elmegreen}{Elmegreen \&
  Elmegreen}{2005}]{Elmegreen2005}
Elmegreen B.~G.,  Elmegreen D.~M.,  2005, ApJ, 627, 632

\bibitem[\protect\citeauthoryear{Fang, Faber, Koo  \& Dekel}{Fang
  et~al.}{2013}]{Fang2013a}
Fang J.~J.,  Faber S.~M.,  Koo D.~C.,   Dekel A.,  2013, ApJ, 776, 63

\bibitem[\protect\citeauthoryear{Forrest et~al.,}{Forrest
  et~al.}{2016}]{Forrest2016}
Forrest B.,  et~al., 2016, ApJ, 818, L26

\bibitem[\protect\citeauthoryear{Gadotti}{Gadotti}{2009}]{Gadotti2009}
Gadotti D.~A.,  2009, MNRAS, 393, 1531

\bibitem[\protect\citeauthoryear{Galametz et~al.,}{Galametz
  et~al.}{2013}]{Galametz2013}
Galametz A.,  et~al., 2013, ApJS, 206, 10

\bibitem[\protect\citeauthoryear{Genzel et~al.,}{Genzel
  et~al.}{2013}]{Genzel2013}
Genzel R.,  et~al., 2013, ApJ, 773, 68

\bibitem[\protect\citeauthoryear{Gonz{\'{a}}lez~Delgado
  et~al.,}{Gonz{\'{a}}lez~Delgado et~al.}{2015}]{Delgado2015}
Gonz{\'{a}}lez~Delgado R.~M.,  et~al., 2015, A{\&}A, 581, A103

\bibitem[\protect\citeauthoryear{Gonz{\'{a}}lez~Delgado
  et~al.,}{Gonz{\'{a}}lez~Delgado et~al.}{2016}]{Delgado2016}
Gonz{\'{a}}lez~Delgado R.~M.,  et~al., 2016, A{\&}A, 590, A44

\bibitem[\protect\citeauthoryear{Grogin et~al.,}{Grogin
  et~al.}{2011}]{Grogin2011}
Grogin N.~A.,  et~al., 2011, ApJS, 197, 35

\bibitem[\protect\citeauthoryear{Guo, Giavalisco, Ferguson, Cassata  \&
  Koekemoer}{Guo et~al.}{2012}]{Guo2012}
Guo Y.,  Giavalisco M.,  Ferguson H.~C.,  Cassata P.,   Koekemoer A.~M.,  2012,
  ApJ, 757, 120

\bibitem[\protect\citeauthoryear{Guo et~al.,}{Guo et~al.}{2013}]{Guo2013}
Guo Y.,  et~al., 2013, ApJS, 207, 24

\bibitem[\protect\citeauthoryear{Guo et~al.,}{Guo et~al.}{2015}]{Guo2015}
Guo Y.,  et~al., 2015, ApJ, 800, 39

\bibitem[\protect\citeauthoryear{Jung et~al.,}{Jung et~al.}{2017}]{Jung2017}
Jung I.,  et~al., 2017, ApJ, 834, 81

\bibitem[\protect\citeauthoryear{Kassin et~al.,}{Kassin
  et~al.}{2012}]{Kassin2012}
Kassin S.~A.,  et~al., 2012, ApJ, 758, 106

\bibitem[\protect\citeauthoryear{Kennicutt}{Kennicutt}{1992}]{Kennicutt1992}
Kennicutt R.~C.,  1992, ApJ, 388, 310

\bibitem[\protect\citeauthoryear{Kennicutt \& Evans}{Kennicutt \&
  Evans}{2012}]{KennicuttJr2012}
Kennicutt R.~C.,  Evans N.~J.,  2012, ARA{\&}A, 50, 531

\bibitem[\protect\citeauthoryear{Koekemoer et~al.,}{Koekemoer
  et~al.}{2011}]{Koekemoer2011}
Koekemoer A.~M.,  et~al., 2011, ApJS, 197, 36

\bibitem[\protect\citeauthoryear{Kriek, van Dokkum, Labb{\'{e}}, Franx,
  Illingworth, Marchesini  \& Quadri}{Kriek et~al.}{2009}]{Kriek2009}
Kriek M.,  van Dokkum P.~G.,  Labb{\'{e}} I.,  Franx M.,  Illingworth G.~D.,
  Marchesini D.,   Quadri R.~F.,  2009, ApJ, 700, 221

\bibitem[\protect\citeauthoryear{Labb{\'{e}} et~al.,}{Labb{\'{e}}
  et~al.}{2005}]{Labbe2005}
Labb{\'{e}} I.,  et~al., 2005, ApJ, 624, L81

\bibitem[\protect\citeauthoryear{Laidler et~al.,}{Laidler
  et~al.}{2007}]{Laidler2007}
Laidler V.~G.,  et~al., 2007, PASP, 119, 1325

\bibitem[\protect\citeauthoryear{Leethochawalit, Jones, Ellis, Stark, Richard,
  Zitrin  \& Auger}{Leethochawalit et~al.}{2016}]{Leethochawalit2016}
Leethochawalit N.,  Jones T.~A.,  Ellis R.~S.,  Stark D.~P.,  Richard J.,
  Zitrin A.,   Auger M.,  2016, ApJ, 820, 84

\bibitem[\protect\citeauthoryear{Liu et~al.,}{Liu et~al.}{2013}]{liu2013}
Liu G.,  et~al., 2013, ApJ, 778, L41

\bibitem[\protect\citeauthoryear{Liu et~al.,}{Liu et~al.}{2016}]{Liu2016}
Liu F.~S.,  et~al., 2016, ApJ, 822, L25

\bibitem[\protect\citeauthoryear{Lotz, Jonsson, Cox  \& Primack}{Lotz
  et~al.}{2010}]{Lotz2010}
Lotz J.~M.,  Jonsson P.,  Cox T.~J.,   Primack J.~R.,  2010, MNRAS, 404, 590

\bibitem[\protect\citeauthoryear{Meurer, Heckman, Leitherer, Kinney, Robert  \&
  Garnett}{Meurer et~al.}{1995}]{Meurer1995}
Meurer G.~R.,  Heckman T.~M.,  Leitherer C.,  Kinney A.,  Robert C.,   Garnett
  D.~R.,  1995, AJ, 110, 2665

\bibitem[\protect\citeauthoryear{Meurer, Heckman, Lehnert, Leitherer  \&
  Lowenthal}{Meurer et~al.}{1997}]{Meurer1997}
Meurer G.~R.,  Heckman T.~M.,  Lehnert M.~D.,  Leitherer C.,   Lowenthal J.,
  1997, AJ, 114, 54

\bibitem[\protect\citeauthoryear{Meurer, Heckman  \& Calzetti}{Meurer
  et~al.}{1999}]{Meurer1999}
Meurer G.~R.,  Heckman T.~M.,   Calzetti D.,  1999, ApJ, 521, 64

\bibitem[\protect\citeauthoryear{Momcheva et~al.,}{Momcheva
  et~al.}{2016}]{Momcheva2016}
Momcheva I.~G.,  et~al., 2016, ApJ, 225, L27

\bibitem[\protect\citeauthoryear{Morris et~al.,}{Morris
  et~al.}{2015}]{Morris2015}
Morris A.~M.,  et~al., 2015, ApJ, 149, 178

\bibitem[\protect\citeauthoryear{Murphy, Chary, Dickinson, Pope, Frayer  \&
  Lin}{Murphy et~al.}{2011}]{Murphy2011}
Murphy E.~J.,  Chary R.-R.,  Dickinson M.,  Pope A.,  Frayer D.~T.,   Lin L.,
  2011, ApJ, 732, 126

\bibitem[\protect\citeauthoryear{Muzzin et~al.,}{Muzzin
  et~al.}{2013}]{Muzzin2013}
Muzzin A.,  et~al., 2013, ApJS, 206, 8

\bibitem[\protect\citeauthoryear{Nelson et~al.,}{Nelson
  et~al.}{2016a}]{Nelson2016}
Nelson E.~J.,  et~al., 2016a, ApJ, 817, L9

\bibitem[\protect\citeauthoryear{Nelson et~al.,}{Nelson
  et~al.}{2016b}]{Nelson2016a}
Nelson E.~J.,  et~al., 2016b, ApJ, 828, 27

\bibitem[\protect\citeauthoryear{Noeske et~al.,}{Noeske
  et~al.}{2007}]{Noeske2007}
Noeske K.~G.,  et~al., 2007, ApJ, 660, L43

\bibitem[\protect\citeauthoryear{Nordon et~al.,}{Nordon
  et~al.}{2013}]{Nordon2013}
Nordon R.,  et~al., 2013, ApJ, 762, 125

\bibitem[\protect\citeauthoryear{Pannella et~al.,}{Pannella
  et~al.}{2009}]{Pannella2009}
Pannella M.,  et~al., 2009, ApJ, 701, 787

\bibitem[\protect\citeauthoryear{Papovich et~al.,}{Papovich
  et~al.}{2015}]{Papovich2015}
Papovich C.,  et~al., 2015, ApJ, 803, 26

\bibitem[\protect\citeauthoryear{Patel, Kelson, Holden, Franx  \&
  Illingworth}{Patel et~al.}{2011}]{Patel2011b}
Patel S.~G.,  Kelson D.~D.,  Holden B.~P.,  Franx M.,   Illingworth G.~D.,
  2011, ApJ, 735, 53

\bibitem[\protect\citeauthoryear{Patel et~al.,}{Patel
  et~al.}{2013}]{Patel2013a}
Patel S.~G.,  et~al., 2013, ApJ, 766, 15

\bibitem[\protect\citeauthoryear{Peng, Ho, Impey  \& Rix}{Peng
  et~al.}{2002}]{Peng2002a}
Peng C.~Y.,  Ho L.~C.,  Impey C.~D.,   Rix H.-W.,  2002, ApJ, 124, 266

\bibitem[\protect\citeauthoryear{Puech, Hammer, Rodrigues, Fouquet, Flores  \&
  Disseau}{Puech et~al.}{2014}]{Puech2014}
Puech M.,  Hammer F.,  Rodrigues M.,  Fouquet S.,  Flores H.,   Disseau K.,
  2014, MNRAS, 443, L49

\bibitem[\protect\citeauthoryear{Rachford et~al.,}{Rachford
  et~al.}{2009}]{Rachford2009}
Rachford B.~L.,  et~al., 2009, ApJS, 180, 125

\bibitem[\protect\citeauthoryear{Radburn-Smith et~al.,}{Radburn-Smith
  et~al.}{2012}]{Radburn-Smith2012}
Radburn-Smith D.~J.,  et~al., 2012, ApJ, 753, 138

\bibitem[\protect\citeauthoryear{Reddy, Steidel, Fadda, Yan, Pettini, Shapley,
  Erb  \& Adelberger}{Reddy et~al.}{2006}]{Reddy2006}
Reddy N.~A.,  Steidel C.~C.,  Fadda D.,  Yan L.,  Pettini M.,  Shapley A.~E.,
  Erb D.~K.,   Adelberger K.~L.,  2006, ApJ, 644, 792

\bibitem[\protect\citeauthoryear{Reddy, Erb, Pettini, Steidel  \&
  Shapley}{Reddy et~al.}{2010}]{Reddy2010}
Reddy N.~A.,  Erb D.~K.,  Pettini M.,  Steidel C.~C.,   Shapley A.~E.,  2010,
  ApJ, 712, 1070

\bibitem[\protect\citeauthoryear{Reddy, Pettini, Steidel, Shapley, Erb  \&
  Law}{Reddy et~al.}{2012}]{Reddy2012}
Reddy N.~A.,  Pettini M.,  Steidel C.~C.,  Shapley A.~E.,  Erb D.~K.,   Law
  D.~R.,  2012, ApJ, 754, 25

\bibitem[\protect\citeauthoryear{Rodr{\'{i}}guez-Puebla, Primack, Behroozi  \&
  Faber}{Rodr{\'{i}}guez-Puebla et~al.}{2016}]{Rodriguez-Puebla2016}
Rodr{\'{i}}guez-Puebla A.,  Primack J.~R.,  Behroozi P.,   Faber S.~M.,  2016,
  MNRAS, 455, 2592

\bibitem[\protect\citeauthoryear{Ro{\v{s}}kar, Debattista, Stinson, Quinn,
  Kaufmann  \& Wadsley}{Ro{\v{s}}kar et~al.}{2008}]{Roskar2008}
Ro{\v{s}}kar R.,  Debattista V.~P.,  Stinson G.~S.,  Quinn T.~R.,  Kaufmann T.,
    Wadsley J.,  2008, ApJ, 675, L65

\bibitem[\protect\citeauthoryear{Salmon et~al.,}{Salmon
  et~al.}{2016}]{Salmon2016}
Salmon B.,  et~al., 2016, ApJ, 827, 20

\bibitem[\protect\citeauthoryear{Sandstrom et~al.,}{Sandstrom
  et~al.}{2013}]{Sandstrom2013}
Sandstrom K.~M.,  et~al., 2013, ApJ, 777, 5

\bibitem[\protect\citeauthoryear{Santini et~al.,}{Santini
  et~al.}{2015}]{Santini2015}
Santini P.,  et~al., 2015, ApJ, 801, 97

\bibitem[\protect\citeauthoryear{Shibuya, Ouchi  \& Harikane}{Shibuya
  et~al.}{2015}]{Shibuya2015}
Shibuya T.,  Ouchi M.,   Harikane Y.,  2015, ApJS, 219, 15

\bibitem[\protect\citeauthoryear{Simons et~al.,}{Simons
  et~al.}{2016}]{Simons2016}
Simons R.~C.,  et~al., 2016, ApJ, 830, 14

\bibitem[\protect\citeauthoryear{Speagle, Steinhardt, Capak  \&
  Silverman}{Speagle et~al.}{2014}]{Speagle2014}
Speagle J.~S.,  Steinhardt C.~L.,  Capak P.~L.,   Silverman J.~D.,  2014, ApJS,
  214, 15

\bibitem[\protect\citeauthoryear{Stott et~al.,}{Stott et~al.}{2014}]{Stott2014}
Stott J.~P.,  et~al., 2014, MNRAS, 443, 2695

\bibitem[\protect\citeauthoryear{Szomoru et~al.,}{Szomoru
  et~al.}{2010}]{Szomoru2010}
Szomoru D.,  et~al., 2010, ApJ, 735, L22

\bibitem[\protect\citeauthoryear{Tacchella et~al.,}{Tacchella
  et~al.}{2015}]{Tacchella2015}
Tacchella S.,  et~al., 2015, Science, 348, 314

\bibitem[\protect\citeauthoryear{Tacchella, Dekel, Carollo, Ceverino, DeGraf,
  Lapiner, Mandelker  \& Primack}{Tacchella et~al.}{2016}]{Tacchella2016a}
Tacchella S.,  Dekel A.,  Carollo C.~M.,  Ceverino D.,  DeGraf C.,  Lapiner S.,
   Mandelker N.,   Primack J.~R.,  2016, MNRAS, 458, 242

\bibitem[\protect\citeauthoryear{Tacchella et~al.,}{Tacchella
  et~al.}{2017}]{Tacchella2017}
Tacchella S.,  et~al., 2017, preprint (arXiv:1704.00733)

\bibitem[\protect\citeauthoryear{Tuffs, Popescu, Volk  \& Fischera}{Tuffs
  et~al.}{2004}]{Tuffs2004}
Tuffs R.~J.,  Popescu C.~C.,  Volk H.~J.,   Fischera J.,  2004, A\&A, 419, 821

\bibitem[\protect\citeauthoryear{Welker, Dubois, Devriendt, Pichon, Kaviraj  \&
  Peirani}{Welker et~al.}{2017}]{Welker2017}
Welker C.,  Dubois Y.,  Devriendt J.,  Pichon C.,  Kaviraj S.,   Peirani S.,
  2017, MNRAS, 465, 1241

\bibitem[\protect\citeauthoryear{Whitaker, van Dokkum, Brammer  \&
  Franx}{Whitaker et~al.}{2012}]{Whitaker2012}
Whitaker K.~E.,  van Dokkum P.~G.,  Brammer G.,   Franx M.,  2012, ApJ, 754,
  L29

\bibitem[\protect\citeauthoryear{Wild, Charlot, Brinchmann, Heckman, Vince,
  Pacifici  \& Chevallard}{Wild et~al.}{2011}]{Wild2011}
Wild V.,  Charlot S.,  Brinchmann J.,  Heckman T.,  Vince O.,  Pacifici C.,
  Chevallard J.,  2011, MNRAS, 417, 1760

\bibitem[\protect\citeauthoryear{Williams, Quadri, Franx, van Dokkum  \&
  Labb{\'{e}}}{Williams et~al.}{2009}]{Williams2009}
Williams R.~J.,  Quadri R.~F.,  Franx M.,  van Dokkum P.,   Labb{\'{e}} I.,
  2009, ApJ, 691, 1879

\bibitem[\protect\citeauthoryear{Wuyts et~al.,}{Wuyts et~al.}{2007}]{Wuyts2007}
Wuyts S.,  et~al., 2007, ApJ, 655, 51

\bibitem[\protect\citeauthoryear{Wuyts et~al.,}{Wuyts et~al.}{2011}]{Wuyts2011}
Wuyts S.,  et~al., 2011, ApJ, 742, 96

\bibitem[\protect\citeauthoryear{Wuyts et~al.,}{Wuyts et~al.}{2012}]{Wuyts2012}
Wuyts S.,  et~al., 2012, ApJ, 753, 114

\bibitem[\protect\citeauthoryear{Wuyts et~al.,}{Wuyts et~al.}{2013}]{Wuyts2013}
Wuyts S.,  et~al., 2013, ApJ, 779, 135

\bibitem[\protect\citeauthoryear{Wuyts et~al.,}{Wuyts et~al.}{2016}]{Wuyts2016}
Wuyts E.,  et~al., 2016, ApJ, 827, 74

\bibitem[\protect\citeauthoryear{Zheng et~al.,}{Zheng et~al.}{2015}]{Zheng2015}
Zheng Z.,  et~al., 2015, ApJ, 800, 120

\bibitem[\protect\citeauthoryear{Zolotov et~al.,}{Zolotov
  et~al.}{2015}]{Zolotov2015}
Zolotov A.,  et~al., 2015, MNRAS, 450, 2327

\bibitem[\protect\citeauthoryear{van Dokkum et~al.,}{van Dokkum
  et~al.}{2013}]{VanDokkum2013}
van Dokkum P.~G.,  et~al., 2013, ApJ, 771, L35

\bibitem[\protect\citeauthoryear{van~der Wel et~al.,}{van~der Wel
  et~al.}{2012}]{VanderWel2012}
van~der Wel A.,  et~al., 2012, ApJS, 203, 24

\bibitem[\protect\citeauthoryear{van~der Wel et~al.,}{van~der Wel
  et~al.}{2014a}]{VanderWel2014a}
van~der Wel A.,  et~al., 2014a, ApJ, 788, 28

\bibitem[\protect\citeauthoryear{van~der Wel et~al.,}{van~der Wel
  et~al.}{2014b}]{VanderWel2014}
van~der Wel A.,  et~al., 2014b, ApJ, 792, L6

\makeatother
\end{thebibliography}
	
	\appendix
	
	\section{Empirical rest-frame UVI colours}
	\begin{figure*}
		\centering
		\includegraphics[width=4.8in]{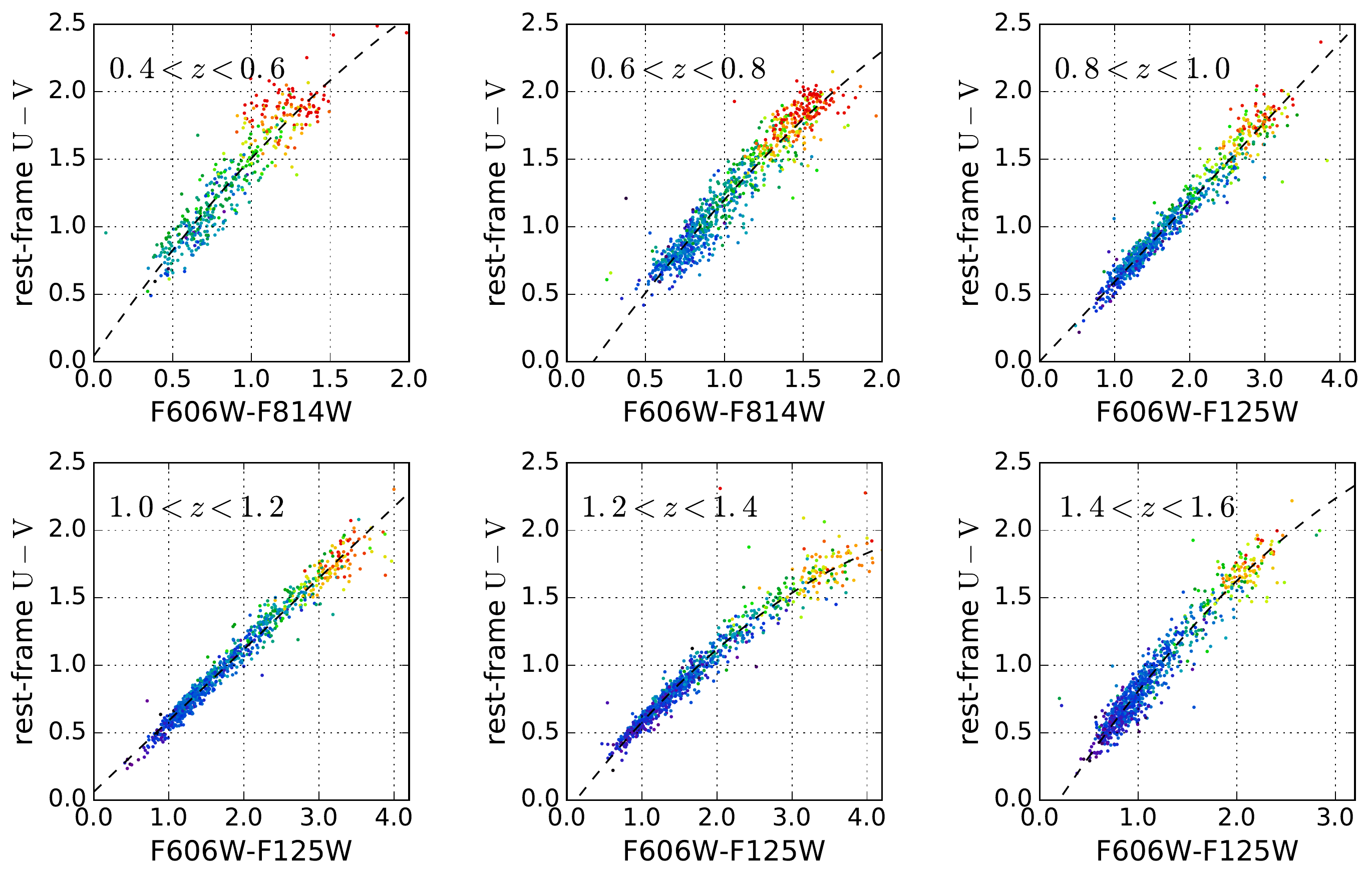}
		\caption{ Fitting the relation between the rest-frame $U-V$ colour and various observer-frame filter combinations.  Each point in the plots represents the integrated colours of one galaxy from the CANDELS photometry catalogue. Dashed curves are best-fitting linear or quadratic models. All the points are colour-coded by sSFR, in the same way as Fig.~\ref{fig:UVI}. These are the relations used in this paper to derive rest-frame $U-V$ colour from \textit{VIJH} aperture photometry. An important criterion in adopting these relations is that red quiescent and red dusty galaxies lie on the same line. That has generally been achieved, as seen by the lack of separation between high and low star-forming galaxies. }
		\label{fig:uvfit}
	\end{figure*}
	\begin{figure*}
		\centering
		\includegraphics[width=4.8in]{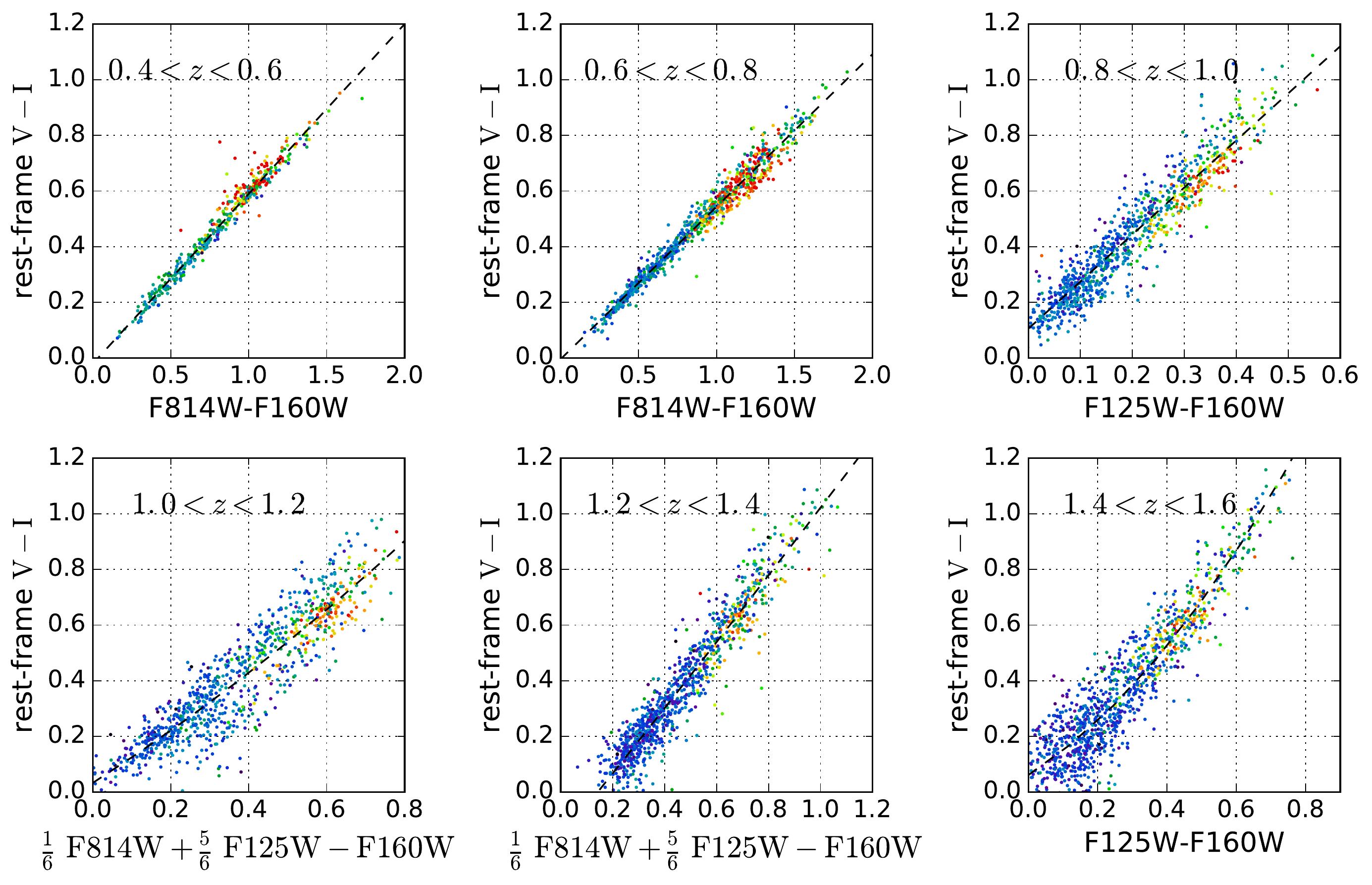}
		\caption{Similar to Fig.~\ref{fig:uvfit}, but fitting the rest-frame $V-I$ colour.  At $1.0<z<1.2$ and $1.2<z<1.4$, empirical combinations of the brightness of three bands, ACS/F814W, WFC3/F125W and WFC3/F160W, are used. These were aimed at by experimentation minimizing the offsets between red quiescent and red dusty galaxies. A further source of scatter appear to come from strong emission lines ($\mathrm{H}\,\alpha$ in F125W at $1.0<z<1.2$) and [\ion{O}{iii}] in F125W at $1.4<z<1.6$. Results from $1.0<z<1.2$ received close inspection, and redshifts beyond $z>1.4$ were deleted as too risky.}
		\label{fig:vifit}
	\end{figure*}
	
	\label{append:A}
	In order to derive the rest-frame colour profiles using multi-band annular photometry, we try to deduce the empirical relations between rest frame colours and observer-frame colours from the CANDELS integrated photometry catalogue \citep{Guo2013, Galametz2013}. For most galaxies in UDS, only  F606W, F814W, F125W, and F160W photometry is available. Therefore in order to make the maximum use of our current data sample, we evaluate the feasibility of using these four filters to constrain rest-frame $U-V$ and $V-I$ colours. 
	
	Fig.~\ref{fig:uvfit} and Fig.~\ref{fig:vifit} show the fitting of the relation between rest-frame colours and observer-frame colours.  Each point in the plots represents the integrated colours of one individual galaxy from the CANDELS photometry catalogue, and the dashed curves are best-fitting models, either as straight lines or quadratic curves. For the rest-frame $U-V$ colour, the typical scatter of our fitting is 0.1--0.2 mag, and no bias about the sSFR is found. For $V-I$, we find the colour scatter increases significantly at high redshift, possibly caused by random photometry uncertainty, but more likely by the increasing equivalent width (EW) of strong emission lines in the observer frame.  The scatter at $0.4<z<0.8$ is smaller than 0.05 dex. However for $0.8<z<1.4$ the scatter can be as large as 0.1 dex, especially in the $1.0<z<1.2$ and $z>1.4$ bins, where the scatter reaches to around 0.2 dex. The non-unit coefficients in the $x$-axis quantity in some panels are the result of empirical efforts to reduce this scatter. We constrain the sample to $z<1.4$ in order to obtain relatively accurate rest-frame photometry and take special caution with the $1.0<z<1.2$ samples by inspecting whether the derived \textit{UVI} colour profiles follow the same evolutionary trend as other redshift bins.  Finally, as shown by \citet{Wuyts2013}, there are no systematic differences between the rest-frame colours derived from the four \textit{VIJH} bands and those derived from the seven \textit{BVizYJH} bands, for the range of rest-frame wavelength and redshifts used here. Further details are given in the figure captions. This indirectly supports the robustness of rest-frame colour deduction in our work.

	\section{A toy model for the composite stellar population}
	\label{sec:uvi_composite}
	
	\begin{figure*}
		\centering
		\includegraphics[width=4.8in]{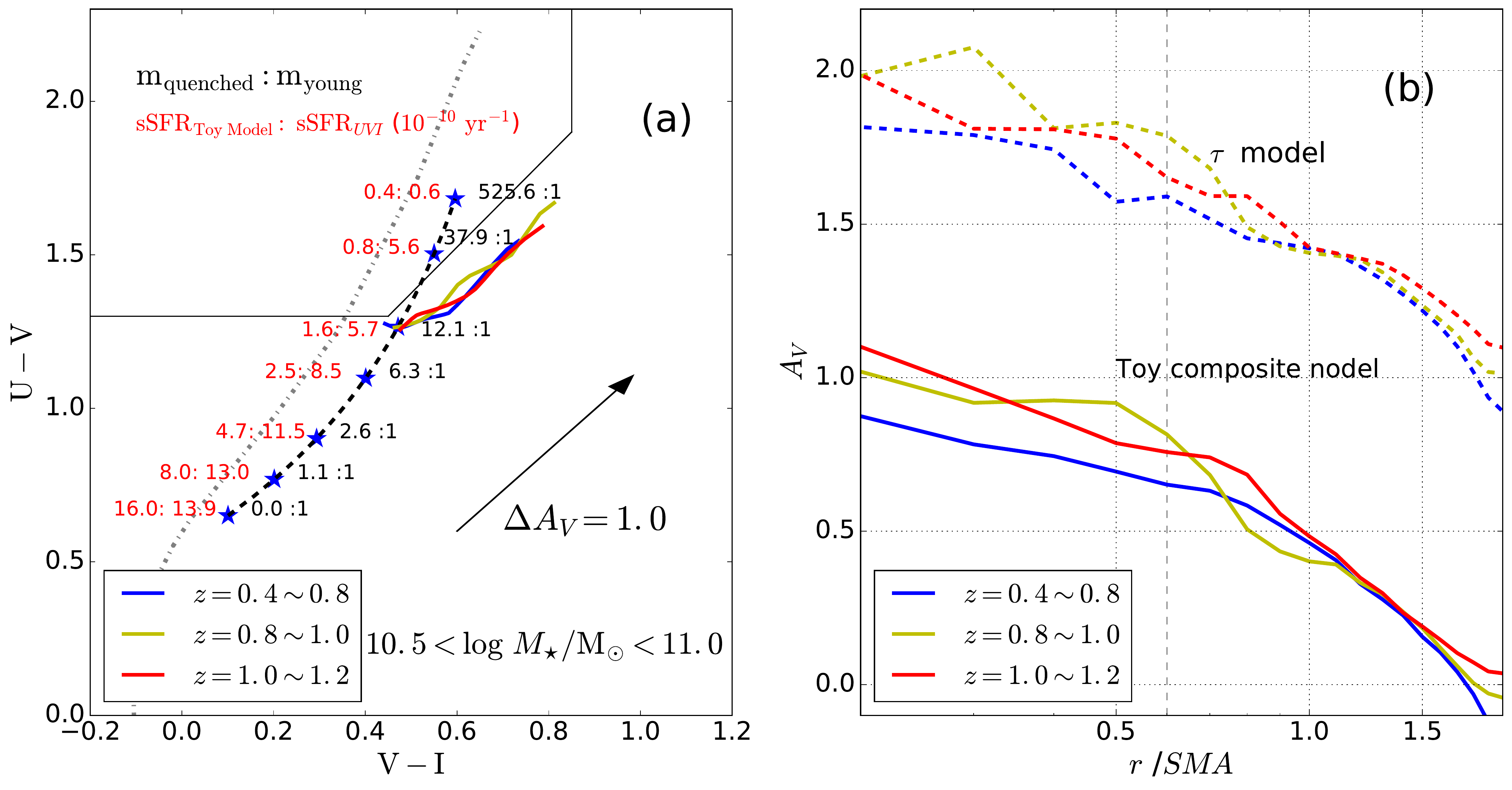}
		\caption{Panel a): A toy composite stellar population model in the 
			\textit{UVI} diagram. The 
			top and bottom stars are typical quenched and blue star-forming 
			galaxies from the CANDELS catalogue; their mass-to-light ratios are 
			CANDELS catalogue values from SED fitting.  These two populations are 
			combined with the mass ratios shown (blue labels), and the resulting \textit{UVI} 
			colours are the intermediate blue stars.  A conventional $\tau$ model with $\tau=1.0$ Gyr 
			plotted in Fig.~\ref{fig:spstrack} is shown again here as the gray curve for comparison.
			The sSFR values deduced from 
			the \textit{UVI} calibration versus the actual values from the toy models are 
			shown as the red labels.  The \textit{UVI} values are fairly accurate near the 
			ends of the range but overestimate sSFR by about $\times$3--6 near the 
			top of the blue cloud and in the green valley.  Compositeness thus 
			appears to preserve the relative rankings of sSFR but compresses the 
			range of deduced  sSFR within the blue cloud. 
			Panel b):  Old versus new $A_V$ profiles for 
			massive galaxies using conventional $\tau$-models (top) as the 
			unreddened starting model versus the composite model (third star down in 
			panel a)).   Using the composite model subtracts roughly 1.0 mag from 
			all $A_V$ values, preserving the absolute $A_V$ gradient but reducing 
			the outer value to zero. }
		\label{fig:composite}
	\end{figure*}
	A toy model  is shown in Fig.~\ref{fig:composite} to illustrate the features of highly composite stellar population. 
	By `composite', we mean strongly bimodal mixtures of old and young stars, which significantly deviate from the predictions of the exponentially declining SFH ($\tau$-model). The integrated populations of galaxies
	may evolve in this direction as bulges quench and star formation continues in the outer parts. Less obvious, though still possible,
	is the compositeness at a given radius, where old and young population may co-exist. In general, the possibility of compositeness is 
	greater at later times, when the oldest stars can be much older than the youngest stars. Hence, compositeness is likely to be more of a concern 
	at lower redshifts.
	
	The top and bottom points at the ends of the black dashed curve in Fig.~\ref{fig:composite} represent two typical
	stellar populations of quenched galaxies and blue star forming galaxies, with little dust extinction. The sSFR and $A_V$
	of the two points are identified from the CANDELS catalogue as described in the main text.  Then we mix the two populations by various mass ratios, as shown by
	$m_{\mathrm{quenched}}:m_{\mathrm{young}}$ in the figure, and locate the mixed populations on the \textit{UVI} diagram by inferring their
	colours from the mass ratio and the mass-light ratios, as retrieved from the CANDELS rest-frame photometry catalogue. These model populations
	are shown as star symbols lying on the black dashed curves in the figure. For each stellar population mixture, we calculate its sSFR from the inferred total mass
	and SFR, indicated as $\mathrm{sSFR}_{\mathrm{Toy\ Model}}$. This is compared with the sSFR that would be derived from the \textit{UVI} calibration method explained
	in Section \ref{sec:UVIcali}, indicated as $\mathrm{sSFR}_{\mathit{UVI}}$.
	
	Finally, three observed colour gradients for the mass bin $\log M_\star/\mathrm{M}_{\sun}=10.5$--11.0 and redshift range $z=0.4$--1.2 are overplotted in Fig.~\ref{fig:composite}, 
	to show as an example how composite models would change our interpolation of sSFR and $A_V$ gradients. 
	We infer the sSFR and $A_V$ profiles by assuming that the colour gradients are caused by
	\emph{uniform} attenuation of the composite stellar population on the dashed curve. The inferred $A_V$ profiles, in comparison with those
	derived from the \textit{UVI} calibration method, are shown in panel b.
	
	The major impact of using the composite model is to greatly reduce the derived value of $A_V$. This occurs because 
	the composite model lies systematically to the right of the $\tau$-laws (see the gray curve), so that deduced $A_V$ is lower.
	For the $A_V$ gradients shown, the composite model subtracts a full 1.0 mag from all $A_V$ values. Relative $A_V$ gradients are 
	preserved, but the outer values are reduced to zero. This exercise shows that compositeness is one approach to reducing outer dust values, 
	if that is deemed necessary. It also illustrates how $\tau$-models tend to maximize $A_V$ values relative to other stellar population possibilities.
	Therefore gas profiles inferred from the composite model would have much larger \emph{relative} radial changing rates due to the zero point shift. Comparing with values given by the composite models, the \textit{UVI} values are fairly accurate near the 
	ends of the black curves, but overestimate sSFR by about $\times$3--4 near the 
	top of the blue cloud and in the green valley.  Compositeness thus 
	appears to preserve the relative rankings of sSFR but compresses the 
	range of deduced  sSFR within the blue cloud. The relatively
	flat trend of colour gradients is not affected because that they are parallel to the Calzetti vector.
	
	Besides this \emph{uniform} dust screen model, we have also tested differential attenuation models (not shown), which mixes dust-free old stars and the dusty star forming population to match the observed $UVI$ colour gradients.  In this case the effective mass-to-light ratio of young stars increases by an additional factor of $\sim$ 4  due to dust attenuation ($A_{V, \mathrm{young star}}\sim 1.5$ mag). As a result, young dusty stars must dominate over the old stars in mass to match the observed colours of galaxy centres, and the high fraction of old stars cannot be achieved as expected. The flatness of sSFR profiles and the shift of $A_V$ profiles are found to be similar with those from the uniform attenuation model. 
	
	In a word, our toy models of composite stellar population shift the $A_V$ profiles downward by the same amount, but cannot alter the relatively flat trend of sSFR profiles, indicating that the active central star-formation described in the main text is robust against the tests.
	
	\bsp	
	\label{lastpage}
\end{document}